%% file: arXiv_two_columns.tex
\documentclass[10pt,journal,cspaper,compsoc]{IEEEtran}

\usepackage[nocompress]{cite}
\usepackage{url,comment,amsthm}
\usepackage{amsmath,amssymb,amsfonts,dsfont}
\usepackage{graphicx}
\usepackage{epsfig,epsf,psfrag,textcomp,tabularx}
\usepackage[usenames,dvipsnames]{xcolor}
\usepackage{subfigure}
\usepackage{float}
\usepackage{stfloats,mathtools}
\usepackage[algoruled,medskip,dontprintsemicolon,linesnumbered,Algorithm]{algorithm}
\usepackage[noend]{algpseudocode}
\usepackage{mathtools}

\newcommand{\indicator}[1]{\mathds{1}_{\left[ {#1} \right] }}
\newcommand{\Fig}[1]{Fig.~\ref{fig:#1}}
\newcommand{\Sec}[1]{Sec.~\ref{sec:#1}}
\newcommand{\Eq}[1]{Eq.~(\ref{eq:#1})}
\newcommand{\Alg}[1]{Alg.~\ref{alg:#1}}
\newcommand{\Line}[1]{Line~\ref{line:#1}}
\newcommand{\Lemma}[1]{Lemma~\ref{lem:#1}}

\newtheorem{property}{Property}
\newtheorem{definition}{Definition}
\newtheorem{lemma}{Lemma}
\newcommand{\Prop}[1]{Property~\ref{prop:#1}}

\usepackage{color}

\newcommand{\Oc}{\mathcal{O}}
\newcommand{\Bc}{\mathcal{B}}
\newcommand{\Cc}{\mathcal{C}}
\newcommand{\Kc}{\mathcal{K}}

\newcommand{\Tc}{\mathcal{T}}

\begin{document}
\bstctlcite{IEEEexample:BSTcontrol} 

\title{
A Sharing- and Competition-Aware Framework\\
for Cellular Network Evolution Planning
}
\author{Paolo~Di~Francesco,~\IEEEmembership{Student~Member,~IEEE,}
        Francesco~Malandrino, 
        Tim~K.~Forde, 
        Luiz~A.~DaSilva,~\IEEEmembership{Senior~Member,~IEEE}
        \\
        (Invited Paper)
\IEEEcompsocitemizethanks{\IEEEcompsocthanksitem All authors are with CTVR / The Telecommunications Centre, University of Dublin, Trinity College, Ireland.
}
\thanks{This work has been partially funded by the Science Foundation Ireland (SFI) under grants No. 10/IN.1/I3007.}}
\maketitle
\thispagestyle{empty}\pagestyle{plain}

\vspace{-2cm}
\begin{abstract}

Mobile network operators are facing the difficult task of significantly increasing capacity to meet projected
demand while keeping CAPEX and OPEX down. We argue that infrastructure sharing is a key consideration in 
operators' planning of the evolution of their networks, and that such planning can be viewed as a stage in the 
cognitive cycle. In this paper, we present a framework to model this planning process while taking into account 
both the ability to share resources and the constraints imposed by competition regulation (the latter quantified 
using the Herfindahl index). 
Using real-world demand and deployment data, we find that the ability to share infrastructure essentially moves 
capacity from rural, sparsely populated areas (where some of the current infrastructure can be decommissioned) to 
urban ones (where most of the next-generation base stations would be deployed), with significant increases in 
resource efficiency.
Tight competition regulation somewhat limits the ability to share but does not entirely 
jeopardize those gains, while having the secondary effect of encouraging the wider deployment of next-generation 
technologies.

\end{abstract}

\begin{keywords}
Cellular networks, Network sharing, Network planning, Real-data
\end{keywords}

\section{Introduction}
\label{sec:intro}

With cellular coverage reaching virtually every human being on the planet, the challenge now lies in
{\em evolving} the existing infrastructure to cope with foreseen explosion in the 
demand for capacity~\cite{cisco}.
Evolution will mean different things at different locations: some parts of the infrastructure will be replaced 
with new-generation equipment, e.g., LTE and its successors; others will be upgraded for the purpose of enhancing 
capacity, e.g., by increasing sectorization; finally, underutilized base stations will be decommissioned, 
possibly permanently.

\begin{figure}[t!]
\psfrag{A}[][][1]{$B$}
\psfrag{B}[][][1]{$A$}
\psfrag{Z}[][][1]{$\Omega$}
\centering
	\includegraphics[width=.45\textwidth]{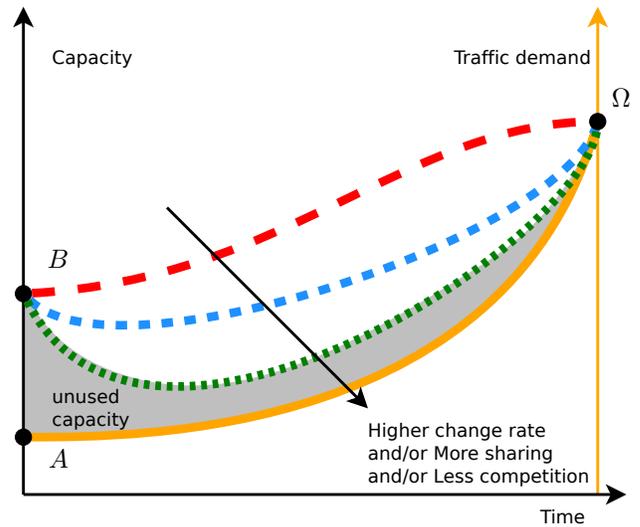}
	\caption{The present and future evolution of cellular infrastructures. The solid yellow line represents the traffic demand, growing from its present value in~$A$ to a much higher one in~$\Omega$. Dashed lines represent different evolutions of network capacity.
	Its present value (in~$B$) is higher than the current demand~$A$, but lower than the future demand~$\Omega$.
	The area between the capacity and demand curves corresponds to unused capacity (shaded in gray).
	Network capacity therefore has to grow from~$B$ to~$\Omega$ in the long term, possibly decreasing in the short term in order to reduce the amount of unused capacity.
	}
	\label{fig:concept}
\end{figure}

The yellow, solid curve in \Fig{concept} represents the network load and its familiar predicted
almost-exponential growth from the current level in~$A$ to a future one in~$\Omega$.
Dashed lines represent possible evolutions of the network capacity:
there is no question that capacity has to increase from its current level at point~$B$ to~$\Omega$,
so as to serve all the demand; in this paper we investigate {\em how} and {\em when} the required changes to the network shall be performed,
so as to efficiently match available capacity to demand, minimizing over-provisioning.

The reason why this matters is the gray area in \Fig{concept}, representing unused network capacity.
Providing capacity that nobody uses is a waste of bandwidth, resources and, ultimately, money; therefore, it is of paramount importance for operators to keep overprovisioning as low as possible.
Furthermore, \Fig{concept} refers to the network as a whole; the relative positions of points~$A,B,\Omega$
can be different in different parts of the topology~\cite{difrancesco2014}.
Extreme cases include some sparsely populated rural areas, where the current capacity may exceed
not only the current but also the future demand, i.e.,~$B>\Omega$, and very dense urban areas, which may have~$A\approx B$.

Ideally, operators would like their capacity to instantly fall from~$B$ to~$A$
at all locations, and would achieve this
by decommissioning as many base stations as possible.
Then, they would follow the demand curve all the way to~$\Omega$,
by updating their infrastructure as the load increases, always
keeping the unutilized capacity (i.e., the gray area in \Fig{concept}) to zero.

Such an idealized view conflicts with the reality that making any change to a network,
be it deploying new base stations, updating or decommissioning existing ones,
requires equipment, work-power, and funds -- all resources that are scarce, and whose usage must be carefully planned.
The number of such changes operators can perform in a given time, e.g., a month, is typically limited, and such a limit directly impacts the speed at which network capacity can go down or up, hence the gray area in \Fig{concept}.

Our first goal is then to study the efficient evolution of the cellular infrastructure in light of
the limited budget of possible network changes.
The input to our problem consists of the current set of base stations, the
(projected, possibly based on real-world measurements and topologies)
future demand and a limited {\em change rate} at which we can update, replace or decommission base stations.
This \emph{change rate} reflects the operators' limitations in terms of how they are able to reshape their
own network.
The output we seek is a list of the changes to perform to the network,
and the time at which to enact each of them.
The overall objective is to keep unused capacity, i.e., the gray area in \Fig{concept}, at a minimum.

In studying this problem we account for two important real-world issues: network {\em sharing} and competition {\em regulation}.
Both are widely studied in the literature, but their impact on the evolution of cellular networks has received
relatively little attention so far.

Active network sharing~\cite{leng2014,doyle2014} refers to roaming-like agreements between mobile operators, where users of each operator are served through both networks indifferently.
It is emerging as a promising way to achieve cost savings and enhanced performance; indeed, running their 
networks in such a shared fashion makes it easier for operators to identify underutilized base stations to 
decommission, as well as making the most out of updated, more highly performing infrastructure.

Competition regulation, on the other hand, has the potential to limit the operators' ability
to reduce unused capacity.
Unused capacity is, from the viewpoint of market regulators, capacity that can be sold to new market players,
typically virtual operators, which would increase the level of competition.
It is entirely plausible, therefore, that existing operators be mandated to keep a certain amount of unused 
spectrum in at least some part of the topology; indeed, a similar condition was recently imposed to O2 and Three 
when they merged their Irish branches~\cite{vodafone-merger}.
Studying how sharing and competition regulation shape the evolution of networks is thus an important contribution 
of this paper.

The algorithms we present are most readily separated into three phases: meeting
demand, regulation compliance, and cost reduction. Each of the phases occur in
series to update the network of an operator in order to provide service to
increasing demand while minimising over-provisioning.

In our view, each individual phase conforms to a instantiation of the cognition
loop. Specifically, each phase of the operation has observation, decision, and
action steps. During observation the current situation is assessed in terms of
the current network, the current demand, the already planned updates, and the
expected demand. Decision involves the application of this situational awareness
to some optimization. Actions take the form of changes to the schedule of
network updates. Note that none of the phases explicitly involves learning.
Rather, each phase implements cognition as optimization based on situational
awareness of network and subscriber state.

Furthermore, the collection of all the phases together provides a more nuanced form of
cognition. This cognition uses understanding of current network infrastructure
to plan future deployments based on the input of expected demand and regulatory
policy. As a unit the three phases of our algorithms periodically receive an
observation of projected demand, whereupon a plan for network updates is
constructed. This plan is then used to decide which base station updates should
be applied to the current infrastructure and the action of making these
adjustments is taken.

The remainder of this paper is organized as follows. 
We present our system model in \Sec{sysmodel}.
In \Sec{problem}, we state and solve the problem of scheduling the changes to our network, and discuss its computational complexity and performance in \Sec{discussion}.
\Sec{scenario} presents our reference scenario, which we use to obtain the results we present in \Sec{results}.
Finally, after a related work review in \Sec{related}, \Sec{conclusion} concludes the paper.

\section{System model}
\label{sec:sysmodel}

Our system model revolves around two main elements: {\em base stations} and {\em subscriber clusters}.

\noindent{\bf Model elements}
Base stations~$b\in\Bc$ are elements of the infrastructure with
a certain position, capacity, and coverage area.
Subscriber clusters~$c\in\Cc$ can correspond to one or more actual users, which can be viewed as co-located. They have a
known position and traffic demand.
We also have operators~$o\in\Oc$, and time periods~$k\in\Kc=\{1\dots K\}$.

\noindent{\bf Base station type}
Base stations are not all equal. They can differ in technology, e.g., GSM, 3G, or LTE; furthermore,
even base stations with the same technology differ in such aspects as frequency of operation and sectorization.
Indeed, upgrading a cellular network essentially means changing their type, e.g., from 3G to LTE.
Even disabling a base station can be seen as changing its type to ``off''.

In our model, possible base station types are collected in set~$\Tc$, and every base station~$b\in\Bc$
has a type~$T(b)\in\Tc$. Decommissioned base stations have the special type~$t_\emptyset$. The type of a base
station determines its coverage and performance, as shown next, as well as its associated cost.

\noindent{\bf Coverage and demand}
Our coverage information comes in the form of binary flags~$\gamma(b,c,t)\in\{0,1\}$, expressing whether
base station~$b$ covers subscriber cluster~$c$
if~$b$ is of type~$t$, i.e., if~$T(b)=t$.
Notice how these values do not depend upon time. We indicate
with~$\delta(b,c)\in\mathds{R}_{+}$ the distance between base stations and subscriber clusters.

\noindent{\bf Requested and served traffic}
For each subscriber cluster~$c$, operator~$o$ and time period~$k$,
we know the traffic demand~$\tau(c,k,o)$ from users of operator~$o$ in cluster~$c$ at period~$k$.
To streamline the notation, we will often write~$\tau(c,k)=\sum_{o\in\Oc}\tau(c,k,o)$.

We also indicate with~$\sigma(b,c,k,o,t)$
the traffic demand that can be met by base station~$b$, of type~$t$,
when servicing users of operator~$o$ in cluster~$c$ at period~$k$.
Similarly to~$\sigma$, we will often drop indices to streamline the notation, and write, e.g.,~$\sigma(c,k,o) = \sum_{b\in\Bc}\sum_{t\in\Tc} \sigma(b,c,k,o,t)$.

\begin{figure}[t!]
\psfrag{b}[c][m]{Base}
\psfrag{s}[c][m]{station}
\psfrag{b1}[c][m]{$b_1$}
\psfrag{b2}[c][m]{$b_2$}
\psfrag{b3}[c][m]{$b_3$}
\psfrag{k1}[c][m]{$k=1$}
\psfrag{k2}[c][m]{$k=2$}
\psfrag{k3}[c][m]{$k=3$}
\psfrag{k4}[c][m]{$k=4$}
\psfrag{k5}[c][m]{$k=5$}
\psfrag{u12}[c][m]{\footnotesize{$x(b_1,2,$}}
\psfrag{u14}[c][m]{\footnotesize{$x(b_1,4,$}}
\psfrag{u21}[c][m]{\footnotesize{$x(b_2,1,$}}
\psfrag{w33}[c][m]{\footnotesize{$x(b_3,3,$}}
\psfrag{x1}[c][m]{\footnotesize{$t_2)=1$}}
\psfrag{x2}[c][m]{\footnotesize{$t_3)=1$}}
\psfrag{x3}[c][m]{\footnotesize{$t_2)=1$}}
\psfrag{x4}[c][m]{\footnotesize{$t_\emptyset$)=1}}
\centering
	\includegraphics[width=.45\textwidth]{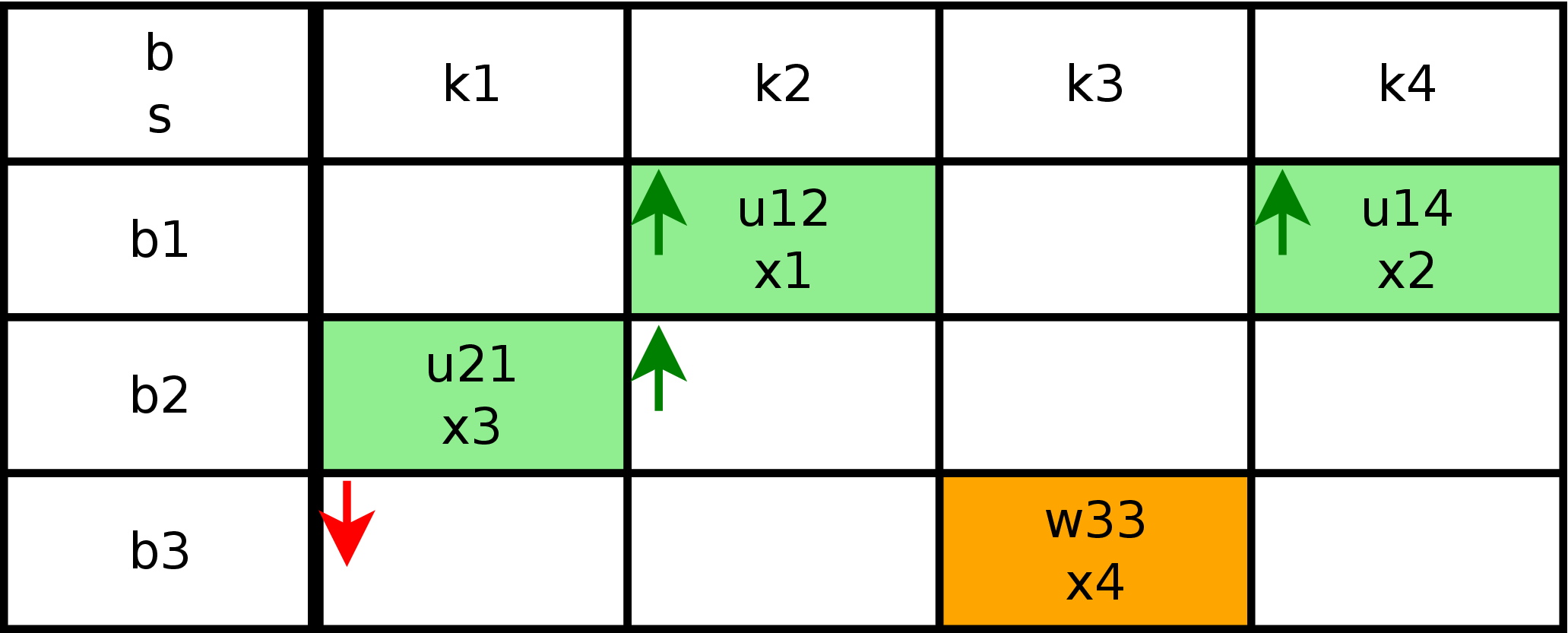}
	\caption{
	An example schedule, with~$|\Bc|=3$ base stations and~$|\Kc|=4$ time periods. The maximum change rate is~$N=1$,
	i.e., we can make at most one change (updating or disabling a base station) per time period.
	Let us assume~$\Tc=\left\{t_\emptyset,t_1,t_2,t_3\right\}$, with~$t_1\dots t_3$ having increasing capacity and the
	same coverage. All base stations start with type~$t_1$.
	Green arrows mark the periods at which base stations need to be updated due to an increased load, i.e., because~$\tau>\sigma$;
	some stations, such as~$b_1$, may need more than one update.
	We update~$b_1$ twice, from~$t_1$ to~$t_2$ and then to~$t_3$, setting~$x(b_1,2,t_2)=x(b_1,4,t_3)=1$.
	Base station~$b_2$ needs an update within~$k=2$; However, we cannot set~$x(b_2,2,t_2)=1$, because doing so
	would violate constraint \Eq{max-n}. This forces us to anticipate the update to~$k=1$, i.e., set~$x(b_2,1,t_2)=1$. 
	Similarly, the red arrow tells us that we would be able to decommission~$b_3$ as soon as~$k=1$,
	but the updates we already scheduled force us to delay until~$k=3$, and set~$x(b_3,3,t_\emptyset)=1$.
	\label{fig:table}
	}
\end{figure}

\noindent{\bf Cost}
Base stations also have an {\em operational cost}~$\kappa(b,T(b))$. Such a cost is base station- and type-dependent, and models such aspects as maintenance, site rental, and energy consumption.
The cost associated with decommissioned base stations is zero, i.e.,~$\kappa(b,t_\emptyset)=0, \forall b\in\Bc$.

\noindent{\bf Network changes}
All our decisions concern network changes.
At each time period~$k$, we may decide to {\em change} the type of base station~$b$, either to a better-performing
type~$t_{dest}$, in order to increase its capacity, or to~$t_\emptyset$ to save on costs, as shown in \Fig{table}.
We track type changes through binary variables:
\begin{equation}
\nonumber
x(b,k,t_{dest})\in\{0,1\}.
\end{equation}
Setting~$x(b,k,t_{dest})=1$ means that, at time~$k$, we change the type of base station~$b\in\Bc$ to~$t_{dest}\in\Tc$.
Doing nothing, i.e., never changing~$b$'s type, is represented by having~$x(b,k,t)=0,\forall k,t$.

Also notice that we can change a base station's type multiple times, i.e., it can be\\
that~$\sum_{k\in\Kc,t\in\Tc}x(b,k,t)>1$. We do not explicitly forbid changing the type to~$t_\emptyset$ and then to
some other type, i.e., first disabling and then re-enabling a base station, although it is unlikely to make sense
in practice (and we never observe that behavior in our performance evaluation).

Changing the type of a base station to~$t_\emptyset$ means reducing network capacity and saving money, i.e.,
going down in \Fig{concept}. On the contrary, moving to a better-performing type means being able to serve
more traffic, hence going up in \Fig{concept}.

In both cases, as discussed in~\Sec{intro}, setting more~$x$-values to~$1$ is linked to going up or down in \Fig{concept}
with a higher slope, i.e., being more effective in reducing the gap between requested and provided capacity.

What limits us is the maximum change rate~$N$, defined as the maximum number of changes
we can make to our network at each time period~$k$. The following constraint must hold:

\begin{equation}
\label{eq:max-n}
\sum_{b\in\Bc,t\in\Tc} x(b,k,t) \leq N, \forall k\in\Kc.
\end{equation}
Having \Eq{max-n} in place means two things. First and most obviously, we can take fewer actions, e.g., 
decommission fewer base stations. Furthermore, we may have to move some actions in time, e.g., decommission a base 
station later than we would like to. Both make us less effective in tracking the demand, i.e., imply a larger gray 
area in \Fig{concept}.

\noindent{\bf Time scale}
It is important to understand the time scale at which our model, and the algorithms described later, work. We are
modeling network {\em planning}, and we are concerned with the evolution of our network over a time span of
months or years.
Each time period~$k\in\Kc$ may correspond to several weeks, and the decisions~$x(b,k,t)$
can be mapped, e.g., to equipment orders or to the schedule of infrastructure deployment teams.
Decommissioning or updating a base station is substantially different from
turning it on and off in order to follow daily traffic fluctuation, as envisioned in ``green networking''
solutions~\cite{green-ajmone,green-on,peng2011}. Indeed, as discussed later, the two solutions are orthogonal
and altogether compatible.

Since networks have to be provisioned for peak loads and not average ones, the~$\tau(c,k)$ values express the
{\em worst-case} amount of traffic requested by subscriber cluster~$c$ during the whole duration of time period~$k$ --
e.g., the amount of traffic (in Mbit) that users in~$c$ will need served during the busiest hour of
time period~$k$. Such values typically come from forecasts and projections; from the viewpoint of our model,
they are an input.

It is also worthwhile to observe that our network must be able to operate even if all the ``worst hours'' of all
subscriber clusters take place at the same time. In other words, while it is possible, and indeed advisable,
to {\em operate} the network so as to take advantage of the low space and traffic correlation in traffic
demand~\cite{difrancesco2014}, such an effect cannot be depended upon in the {\em planning} phase.

\begin{figure}[t!]
\psfrag{A}[c][m]{\textsf{\Alg{capacity}}}
\psfrag{B}[c][m]{\textsf{\Alg{competition}}}
\psfrag{C}[c][m]{\textsf{\Alg{disable}}}
\psfrag{S}[c][m]{\textsf{Performance assessment}}
\psfrag{t}[c][m]{\footnotesize{\textsf{(Simulators, real data...)}}}
\psfrag{y}[l]{\footnotesize{$\sigma$}}
\psfrag{z}[c]{\footnotesize{$\Cc,\Bc,\tau\dots$}}
\psfrag{x}[r]{\footnotesize{$x$}}
\centering
	\includegraphics[width=.45\textwidth]{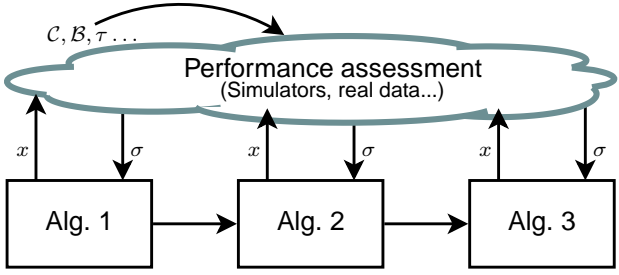}
	\caption{
	Assessing the performance within our model and solution concept.
	Network performance depends on topology, traffic demand
	and decisions; however, our model does not explicitly represent
	such a dependency.
	Our algorithms rely instead on an external block, represented by the cloud in the figure. Given as an input the network topology
	(base stations~$\Bc$, subscriber clusters~$\Cc$), demand~$\delta$, and decisions~$x$, it returns
	the traffic that can be served by the network,~$\sigma$, as an output.
	The most straightforward way of implementing such a block is network simulation;
	however, other approaches are possible. As discussed in \Sec{scenario}, we implement
	the performance assessment block leveraging real-world traces.
	}
	\label{fig:simulator}
\end{figure}

\noindent{\bf Assessing network performance}
It is important to remark that our model does not explicitly include a representation of how the served traffic~$\sigma$ depends
on the other parameters and variables, e.g., our decisions~$x$.
As we see in \Fig{simulator}, the~$\sigma$-values are obtained through an external performance assessment block.
In addition to keeping our model simple, this
choice affords us a higher degree of flexibility: we can interface our model with a simulation tool, or leverage any real-world data available to us, as discussed in \Sec{scenario}.

\noindent{\bf Competition}
Healthy competition within the mobile market is of constant concern to regulators, as
dominance on the part of large operators can lead to market abuses. A common regulatory
tool to measure the level of competition and market concentration is the Herfindahl index
(HHI)~\cite{mergers, EULaw}. It is given by the sum of the squares of shares held by each
operator in the market, and takes values between 0 (a multitude of operators with a zero-share)
and 1 (a monopolist with a 100\% share).
The HHI can be used to assess concentration in different aspects of the market, e.g.,
overall market share, concentration in ownership of spectrum and concentration
in ownership of network infrastructure.

In our scenario, we need to define a {\em local} version of HHI, specific to each
subscriber cluster (as well as to each time period). Furthermore, we have to account not
only for the operators currently in~$\Oc$, but also for new operators that may enter the
market if the conditions are favorable -- typically mobile virtual operators (MVNOs).
Our version of the HHI is thus given by:
\begin{equation}
\label{eq:hhi}
H(c,k)=
\left(\frac{\sigma(c,k)-\tau(c,k)}{\sigma(c,k)}\right)^2
+
\sum_{o\in\Oc}\left (\frac{\tau(c,k,o)}{\sigma(c,k)}\right)^2
\end{equation}
In the denominator of \Eq{hhi} we always find the total capacity~$\sigma(c,k)$
available to subscriber cluster~$c$ at time period~$k$ (recall our conventions
about dropping indices). In the numerator we have the traffic of current operators
in~$\Oc$ in the summation, and the spare capacity, i.e., the traffic of potential
new operators, in the other term.

When two operators deploy and manage their networks in a shared fashion, they
behave as one from the competition viewpoint. Therefore, the set~$\Oc$ shrinks,
and the HHI in \Eq{hhi} increases. In our model, regulators require that the HHI not exceed a value~$H_{\max}$ in at least a significant portion of the topology.

\section{Problem formulation and solution}
\label{sec:problem}

In this section, we address the following problem. Given the future demand~$\tau(c,k,o)$, and the maximum
change rate~$N$, how should each operator schedule the network
changes, i.e., set the~$x$-variables?
Operators have three goals:

{\bf Goal 1} is meeting the {\em traffic demand}, i.e., having:
\begin{equation}
\label{eq:goal1}
\sigma(c,k,o)\geq\tau(c,k,o),\forall c\in\Cc,k\in\Kc,o\in\Oc.
\end{equation}
\Eq{goal1} says that for all subscriber clusters~$c$, time periods~$k\in\Kc$ and operators~$o\in\Oc$, the
provided capacity~$\sigma$
must equal (or exceed) the demand~$\tau$.

{\bf Goal 2} is complying with existing regulation:
\begin{equation}
\label{eq:goal2}
\sum_{c\in\Cc}\indicator{H(c,k)\leq H_{\max}}\geq \phi \cdot|\Cc|,\forall k\in\Kc.
\end{equation}
\Eq{goal2} imposes that, for each time period, at least a fraction $\phi$ of demand clusters
-- enough for a new operator to start building its network~\cite{vodafone-merger,pakistan} --
have an HHI (as defined in \Eq{hhi}) not exceeding the limit~$H_{\max}$.

{\bf Goal 3} is to minimize costs:
\begin{equation}
\label{eq:goal3}
\min \sum_{b\in\Bc,k\in\Kc} \kappa(b,T(b)).
\end{equation}
An obvious way to decrease the quantity in \Eq{goal3} is setting the type of some base stations to~$t_\emptyset$,
whose associated cost is~$0$, i.e., disabling them.

Multi-objective problems,
where some kind of trade-off between different goals is sought, are in general hard to formulate and harder to solve.
Thankfully, in our case goals have a clear hierarchy: the first two goals must be met through as few
changes to the network as possible; any remaining change can be used to pursue the third goal.
Indeed, the first two goals can be treated as constraints, and the third one is the objective we seek to optimize.

\subsection{Solution concept}

Our aim is to exploit the hierarchy of the goals stated above, as well as their features, to devise a
solution concept that addresses them in sequence.

We begin by defining a class of network changes, that we call {\em capacity-preserving changes}, as follows:
\begin{definition}
Changing the type of a base station~$b$ from~$t_{orig}$ to~$t_{dest}$ is {\em capacity-preserving} if
the capacity available to each subscriber cluster does not decrease, i.e.,
\begin{equation}
\label{eq:preserve}
\pi(b,t_{orig},t_{dest})=1 \Leftrightarrow \sigma(b,c,t_{dest})\geq\sigma(b,c,t_{orig}), \forall c\in\Cc.
\end{equation}
\end{definition}

Intuitively, capacity-preserving network changes increase the capacity available to certain subscriber clusters,
without hurting others. Increasing the number of sectors of a base station is a capacity-preserving change, as is
replacing a GSM base station with an LTE-800 one, having the same coverage and a higher capacity. Replacing the same
GSM base station with an LTE-2600 one is not capacity-preserving, as the new base station will have smaller coverage
and some subscriber clusters, namely the ones covered by the old base station but not by the new one, will suffer a
decrease in their available capacity. Similarly, changing any base station's type
to~$t_\emptyset$ is not capacity-preserving.

We are now in the position of proving the following useful properties:
\begin{property}
\label{prop:updates-enough}
Both goal 1 and goal 2 can be reached through capacity-preserving changes alone, i.e.,
changes that comply with \Eq{preserve}.
\end{property}
\begin{IEEEproof}
See the Appendix.
\end{IEEEproof}
\begin{property}
\label{prop:conflict12}
If the initial configuration satisfies goal 1, then pursuing goal 2 by scheduling further
capacity-preserving changes
does not compromise goal 1.
\end{property}

\begin{figure}[t!]
\psfrag{c1}[c][m]{\scriptsize{Capacity}}
\psfrag{c2}[c][m]{\scriptsize{shortage?}}
\psfrag{c3}[c][m]{\scriptsize{Competition}}
\psfrag{c4}[c][m]{\scriptsize{rule breach?}}
\psfrag{c01}[c][m]{\scriptsize{Coverage,}}
\psfrag{c02}[c][m]{\scriptsize{capacity, and}}
\psfrag{c03}[c][m]{\scriptsize{competition}}
\psfrag{c04}[c][m]{\scriptsize{respected?}}
\psfrag{f1}[c][m]{\scriptsize{Find subscriber}}
\psfrag{f2}[c][m]{\scriptsize{clusters with}}
\psfrag{f3}[c][m]{\scriptsize{problems}}
\psfrag{s2}[c][m]{\scriptsize{base station}}
\psfrag{s01}[c][m]{\scriptsize{Select}}
\psfrag{s02}[c][m]{\scriptsize{most urgent}}
\psfrag{s03}[c][m]{\scriptsize{subscriber cluster}}
\psfrag{f01}[c][m]{\scriptsize{Find suitable}}
\psfrag{f02}[c][m]{\scriptsize{base stations to}}
\psfrag{f03}[c][m]{\scriptsize{solve the problem}}
\psfrag{u01}[c][m]{\scriptsize{Update the most}}
\psfrag{u02}[c][m]{\scriptsize{appropriate one as}}
\psfrag{u03}[c][m]{\scriptsize{late as possible}}
\psfrag{d01}[c][m]{\scriptsize{Decommission as early}}
\psfrag{d02}[c][m]{\scriptsize{as possible}}
\psfrag{n}[c][m]{\small{no}}
\psfrag{y}[c][m]{\small{yes}}
\centering
	\includegraphics[width=\columnwidth]{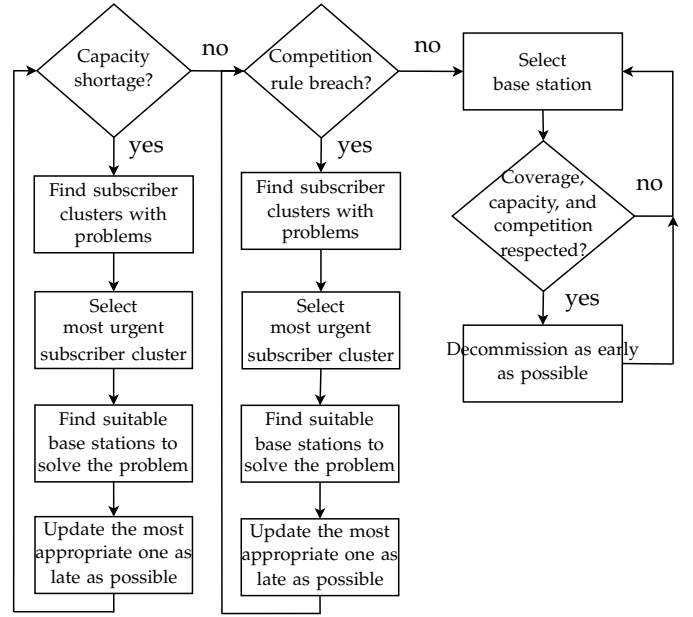}
	\caption{Our solution concept. The three phases correspond to our three objectives -- meeting the traffic
	demand, complying with competition regulation, and reducing costs. Within each phase, we proceed in a similar way: identify
	the subscriber clusters that call for action, e.g., whose demand is not satisfied; identify the base stations that can
	be acted upon to solve the problem,
	i.e., whose type shall be changed, and
	schedule the needed action
	at the most appropriate time. Notice that once we are done with a phase we never come back to it, and subsequent
	decisions are guaranteed not to jeopardize its objective.}
	\label{fig:solution}
\end{figure}

Exploiting these properties, we propose the solution concept
shown in \Fig{solution}, where objectives are
addressed in sequence. Specifically, we first address
goal 1, and do so by
scheduling capacity-preserving network changes (\Prop{updates-enough} guarantees that it is sufficient).
Then, we schedule further capacity-preserving changes in order to reach goal 2: \Prop{updates-enough}
again guarantees that it is possible, and \Prop{conflict12} makes sure that doing so will not jeopardize goal 1.
Finally, we use any remaining changes we can make to the network to pursue goal 3,
as long as doing so does not conflict with goals 1 and 2. Notice that in this last step we are not restricted to capacity-preserving
changes, e.g., we can decommission base stations by changing their type to~$t_\emptyset$.

With reference to \Fig{solution}, we can clearly see how the three goals stated above correspond to three
phases in the algorithm. Within each phase, we proceed in a similar, greedy way: identify the problems and find the most
urgent one to fix; identify the possible actions and find the most appropriate one; and schedule said action at the
most appropriate time.

It is worth stressing that from our viewpoint the future demand, i.e. the $\tau$-values, is but an input to
our problem. In practical settings, such demand will not be known with precision, and this will call for 
appropriate action, e.g., considering a safety margin. Our approach, however, remains unchanged.

\subsection{Individual phases}

\Alg{capacity} summarizes the steps we take in the first phase, where our objective is making sure that
the traffic demand is met at all times and for all subscriber clusters.
It works unmodified with and without network sharing: if there is no sharing, each
operator will run \Alg{capacity} independently, feeding it its own network and its own
load. If operators are performing their updates in a shared fashion,
then \Alg{capacity} will be run only once, on the joint network and the total load.

\begin{algorithm}[t]
\begin{algorithmic}[1]
\Require $\Bc,\Cc,\Kc,\tau$
\While{{\bf true}}
	\State $\sigma \gets \texttt{assess}(x,\delta,\gamma,\tau)$ \label{line:alg1-sim}
	\State $\texttt{problems} \gets \{(c,k)\in\Cc\times\Kc: \sigma(c,k) < \tau(c,k) \}$ \label{line:alg1-findproblem}
	\If{$\texttt{problems}\equiv\emptyset$} \label{line:alg1-checkempty}
		\State{{\bf break}}
	\EndIf
	\State $c^\star,k^\star \gets \arg \min_{(c,k)\in\texttt{problems}} k$ \label{line:alg1-mosturgent}
	\State $ \texttt{actions} \gets \{ (b,t)\in\Bc\times\Tc: \gamma(b,c^\star,t)=1 \wedge \pi(b,T(b),t)=1 \}$ \label{line:alg1-findbss}
	\State $ b^\star \gets \arg \min_{(b,t)\in\texttt{actions}} \delta(b,c^\star)$ \label{line:alg1-bestbs}
	\State $ t^\star \gets \arg \min_{(b,t)\in\texttt{actions}\colon \sigma(c,k,t)\geq\tau(c,k)} \kappa(b,t)$ \label{line:alg1-besttype}
	\State $ \hat{k} \gets \arg \max_{h=0}^{k^\star} \{ h: \sum_{b\in\Bc,t\in\Tc} x(b,k,t) < N \} $ \label{line:alg1-besttime}
	\State $ x(b^\star,\hat{k},t^\star) \gets 1$ \label{line:alg1-schedule}
\EndWhile
\Return $x$
\end{algorithmic}
\caption{Phase 1: ensuring that traffic demand is met.
  \label{alg:capacity}}
\end{algorithm}

The first thing we do is, in \Line{alg1-sim}, to assess the performance we obtain from
currently-scheduled actions, hence obtain the~$\sigma$-values
representing the traffic that can be served for each subscriber cluster.
With reference to \Fig{simulator}, calling function~$\texttt{assess}$
corresponds to entering the ``performance assessment'' cloud.

In \Line{alg1-findproblem}, we look for {\em struggling} subscriber clusters, i.e.,~$(c,k)$ pairs
for which \Eq{goal1} does not hold. If there is no such pair
(\Line{alg1-checkempty}), then we are done and can move to phase 2. Otherwise, we proceed to \Line{alg1-mosturgent}, where we
identify the~$(c^\star,k^\star)$ pair that needs our attention next. In our case, we tackle the issue happening first, i.e., the one
with the lowest~$k^\star$.

So far, we have decided to perform a network change to tackle the capacity shortage affecting subscriber cluster~$c^\star$
at time period~$k^\star$. The set of base stations that we could decide to upgrade is identified in \Line{alg1-findbss},
and corresponds to the set of~$(b,t)$ pairs of base stations~$b$ such that
(i)~$b$ would cover~$c^\star$ if its technology were set to~$t$, and (ii) the change would be capacity-preserving.
Recall that we are relying on \Prop{updates-enough} and \Prop{conflict12} to design the first two steps of our solution concept,
and those properties only hold for capacity-preserving changes.

Among the base stations we may change, we have to identify the most appropriate one~$b^\star$; in
\Line{alg1-bestbs}, we simply select the one closest to~$c^\star$.
In \Line{alg1-besttype}, we select the type~$t^\star$ to update base station~$b^\star$ to. We select, among the types that
would restore the capacity constraint \Eq{goal1}, the one with minimum cost.

Last, we need to schedule the actual upgrade. We want to do so as late as possible, but no later than period~$k^\star$.
Therefore, in \Line{alg1-besttime}, we select the latest period between~$0$ and~$k^\star$, in which we can still do something, i.e.,
for which we have scheduled to change the type of no more than~$N-1$ base stations. Identified such a period~$\hat{k}$, we proceed with
scheduling the upgrade in \Line{alg1-schedule},
by setting the appropriate~$x$-value to~$1$, and move to the next iteration.
In the choice of~$\hat{k}$,
we can clearly see the relationship between the change rate~$N$ and our ability to keep unused capacity (i.e., the gray area
in \Fig{concept}) to a minimum. Setting~$\hat{k}=k^\star$ would mean making the change when needed, hence deploying no unused capacity;
being forced to have~$\hat{k}<k^\star$ means adding some network capacity that will be unused until time~$k^\star$.
As we clearly see from \Line{alg1-besttime}, the likelihood that we have to do so increases as~$N$ gets smaller.

\begin{algorithm}[t]
\begin{algorithmic}[1]
\Require $\Bc,\Cc,\Kc,\tau$
\While{{\bf true}}
	\State $\sigma \gets \texttt{assess}(x,\delta,\gamma,\tau)$ \label{line:alg2-sim}
	\State $\texttt{problems} \gets \{(c,k)\in\Cc\times\Kc: H(c,k) < H_{\max} \}$ \label{line:alg2-findproblem}
	\If{$|\texttt{problems}|\leq (1-\phi)\cdot|\Cc|$} \label{line:alg2-checkempty}
		\State{{\bf break}}
	\EndIf
	\State $c^\star,k^\star \gets \arg \min_{(c,k)\in\texttt{problems}} k$ \label{line:alg2-mosturgent}
	\State $ \texttt{actions} \gets \{ (b,t)\in\Bc\times\Tc: \gamma(b,c^\star,t)=1 \}$ \label{line:alg2-findbss}
	\State $ b^\star \gets \arg \min_{(b,t)\in\texttt{actions}} \delta(b,c^\star)$ \label{line:alg2-bestbs}
	\State $ t^\star \gets \arg \max_{(b,t)\in\texttt{actions}\colon \pi(b,T(b),t)=1} \sigma(b,c^\star,t)$ \label{line:alg2-besttype}
	\State $ \hat{k} \gets \arg \max_{h=0}^{k^\star} \{ h: \sum_{b\in\Bc,t\in\Tc} x(b,k,t) < N \} $ \label{line:alg2-besttime}
	\State $ x(b^\star,\hat{k},t^\star) \gets 1$ \label{line:alg2-schedule}
\EndWhile
\Return $x$
\end{algorithmic}
\caption{Phase 2: enforcing competition constraints.
  \label{alg:competition}}
\end{algorithm}

\Alg{competition} ensures that the competition constraint \Eq{goal2} is met. It has
the same structure as \Alg{capacity}, with some differences worth highlighting.
The problematic subscriber clusters, identified in \Line{alg2-findproblem}, are the
ones where the HHI exceeds the value~$H_{\max}$.
The base station type~$t^\star$ to switch to is selected as the one that offers the highest capacity,
so as to meet the constraint \Eq{goal2} with the smallest number of changes.
Finally, the termination condition (\Line{alg2-checkempty}) is triggered
if at least~70\% of subscriber clusters have a sufficiently low HHI.

\begin{algorithm}[t]
\begin{algorithmic}[1]
\Require $\Bc,\Cc,\Kc,\tau$
\ForAll{$b\in\Bc$} \label{line:alg3-for-b}
 \ForAll{$t\in\Tc\setminus\{T(b)\}$} \label{line:alg3-for-t}
  \State{$\texttt{save}(b,t)\gets\max(0,\kappa(b,T(b))-\kappa(b,t))$} \label{line:alg3-savings}
 \EndFor
\EndFor
\State{$\textbf{sort} \texttt{ save } \textbf{DESC}$} \label{line:alg3-sort}
\ForAll{$(b,t)\in\texttt{save}$}
 \State{$\hat{k} \gets \arg \min_{h=0}^{K} \{ h: \sum_{b\in\Bc,t\in\Tc} x(b,k,t) < N \} $} \label{line:alg3-besttime}
 \State{$x(b,\hat{k},t)\gets 1$} \label{line:alg3-schedule}
 \State $\sigma \gets \texttt{assess}(x,\delta,\gamma,\tau)$ \label{line:alg3-sim}
 \If{\Eq{goal1}~or~\Eq{goal2} do not hold} \label{line:alg3-check}
  \State{$x(b,\hat{k},t)\gets 0$}  \label{line:alg3-revert}
 \EndIf
\EndFor
\Return $x$
\end{algorithmic}
\caption{Phase 3: reducing costs.
  \label{alg:disable}}
\end{algorithm}

After \Alg{capacity} and \Alg{competition}, we are left with a set of capacity-preserving changes to the network
that ensure that goals 1 and 2 (\Eq{goal1} and \Eq{goal2} respectively) are met.
We now seek to schedule further changes, with the objective of minimizing the cost as defined in \Eq{goal3}, i.e.,
attaining goal 3. Notice that we are not relying on \Prop{updates-enough} and \Prop{conflict12} anymore, and are
thus free to schedule non-capacity-preserving changes if need be.

We proceed as shown in \Alg{disable}, and begin by assessing, for each base station~$b$ (\Line{alg3-for-b}) and new type~$t$
(\Line{alg3-for-t}), how much we could save by switching~$b$'s type from~$T(b)$ to~$t$ (\Line{alg3-savings}). Then we examine
the potential changes, starting from the one yielding the most savings (\Line{alg3-sort}), and simply try them out (\Line{alg3-schedule}),
scheduling them at the earliest possible time period, as shown in \Line{alg3-besttime}. In \Line{alg3-sim} we assess the impact of
the newly-scheduled change on the capacity, i.e., the~$\sigma$-values: if either \Eq{goal1} or \Eq{goal2} do not hold, i.e., if
the new change impairs goal 1 or goal 2, we revert it in \Line{alg3-revert}; otherwise, the change is confirmed.

The overall effect of \Alg{disable} is scheduling further network changes, in addition to the ones decided in
\Alg{capacity} and \Alg{competition}, with the purpose of saving money, i.e., reducing the operational costs as defined in
\Eq{goal3}.
These changes are guaranteed not to jeopardize goal 1 and goal 2, thanks to the explicit check in \Line{alg3-check}.
It is worth noting that changes are scheduled, in \Line{alg3-besttime}, as {\em early} as possible -- conversely,
the equivalent lines in \Alg{capacity} and \Alg{competition} seek to schedule changes as late as possible.
Notice that the check in \Line{alg3-check} also implies that all subscriber clusters must be covered
by at least one base station, i.e., no cost-saving action can be taken if that implies shrinking
network coverage. As we will see in \Sec{results}, this implicit constraint has an impact on the
amount of savings we can achieve.

\section{Solution properties}
\label{sec:discussion}

In this section, we examine the success conditions,
computational complexity and optimality of our algorithms.
We prove the
properties for \Alg{capacity}, but the same holds for \Alg{competition} and
\Alg{disable} as well, which have the same structure.

\subsection{Computational complexity}

Our algorithms have been designed with scalability in mind, and exhibit low complexity, linear in the number
of base stations. More formally:
\begin{property}
\label{prop:complexity}
The worst-case, {\em combined} time complexity of all our algorithms is~$O(|\Bc|log|\Bc|)$.
\end{property}
\begin{IEEEproof}
See the Appendix.
\end{IEEEproof}
This result allows us to efficiently tackle large-scale, real-world topologies, as we see in \Sec{results}.
Also notice that \Prop{complexity} refers to the combined complexity of our solution concept, i.e.,
all the algorithms described in \Sec{problem}, and to the worst case;
real-world cases such as the one we consider in \Sec{results} show a substantially lower complexity.

\subsection{Optimality}

In the following, we assess how close our algorithms perform with respect to the optimum. 
Our algorithms make two kinds of decisions: {\em scheduling}, i.e., deciding when to perform network changes, and
{\em choosing} the changes to make. The optimality of these decisions is discussed separately.

We state and prove our properties with reference to \Alg{capacity},
i.e., the first step in \Fig{solution}; however, since the following steps have the same structure, similar arguments hold.

\subsubsection{Scheduling}

The question we look at is the following: given the times~$k^\star$ by which changes need to be applied,
how good are we at picking the time~$\hat{k}$ at which changes are actually performed?

We indicate with~$\mathbf{x}=(x_k)$ the vector of changes we have to schedule, i.e.,~$x_k$ is the number of
base stations whose type has to be changed within time period~$k$.
We begin by proving the following lemma, stating a necessary condition
under which it is possible to schedule a set of changes:
\begin{lemma}
\label{lem:necessary}
The following condition is necessary for a set of updates to be schedulable:
\begin{equation}
\label{eq:necessary}
\sum_{h=1}^{k} x_h\leq kN,\forall k\in\Kc.
\end{equation}
\end{lemma}
\Lemma{necessary} can be verified by inspection of \Eq{necessary}, as the total number of changes made
is upper bounded by the change rate multiplied by the time in which the changes must be made.
\Lemma{necessary} says that any change set not satisfying \Eq{necessary} is impossible to schedule.
Notice that we have not proven that condition \Eq{necessary} is sufficient, i.e., that
if a set of changes does satisfy it then it is possible to schedule it, nor we know how to actually perform the scheduling.
Thankfully, we can prove that \Alg{capacity} does the job:
\begin{property}
\label{prop:feasible}
If a set of changes satisfies \Eq{necessary}, then \Alg{capacity} is able to schedule it.
\end{property}
\begin{IEEEproof}
See the Appendix.
\end{IEEEproof}

\Prop{feasible} says that \Alg{capacity} can schedule all sets of changes satisfying \Eq{necessary}, and \Lemma{necessary}
says that all other sets of changes are impossible to schedule, no matter the algorithm.
It follows that if it is possible to schedule a given set
of changes, then \Alg{capacity} will do it. This is important, but tells us nothing about how good \Alg{capacity} is at
minimizing the unused capacity, i.e., the gray area in \Fig{concept}. Specifically, if~$k^\star_b$ is the time period
at which the type of base station~$b$ needs to be changed and~$\hat{k}_b$ is the time at which the change is performed, we would
like to minimize the quantity:

\begin{equation}
\label{eq:tomin}
\sum_{b\in\Bc}\left( k^\star_b-\hat{k}_b \right ).
\end{equation}
Again, \Alg{capacity} happens to be as effective as it gets:
\begin{property}
\label{prop:optimal}
The schedule returned by \Alg{capacity} minimizes the quantity in \Eq{tomin}.
\end{property}
\begin{IEEEproof}
See the Appendix.
\end{IEEEproof}

\subsubsection{Choice}

After proving \Prop{feasible} and \Prop{optimal}, we may be tempted to conclude that our approach is altogether optimal,
i.e., shrinks the gray area in \Fig{concept} to the absolute minimum. Regrettably, this is not the case: while the scheduling,
i.e., deciding {\em when} making changes to the network, is optimal, we cannot make the same claim about the {\em choice}
of the changes to make, e.g., the base stations
whose type is to be changed.

Indeed, optimally choosing the base stations to change is an NP-hard problem. (We skip the proof, which is based on reduction
from the set-covering problem.) Greedy heuristics such as the one employed in \Alg{capacity} are widely adopted when dealing with
NP-hard problems; indeed, inapproximability results show~\cite{inapprox} that no better solutions than the ones provided by greedy
algorithms exist unless~$P=NP$.
In other words, the best possible polynomial-time approximation for our problem yields a solution that is no closer to the optimum (except for a constant factor) than the one of our algorithms.

\subsection{Summary}

From our discussion, we can conclude that our algorithms exhibit a remarkably low level of complexity, and can schedule
network changes in an optimal way, i.e., keeping the gap between the time when a change is needed and when it is applied to the
minimum.

The choice of such changes is, in general, not optimal. On the other hand, greedy approaches similar to
the one we adopt are commonly used in the literature~\cite{inapprox}, and have been shown to perform remarkably well in practice.

\section{Reference scenario}
\label{sec:scenario}

\begin{figure}
\centering
\includegraphics[width=0.2\textwidth]{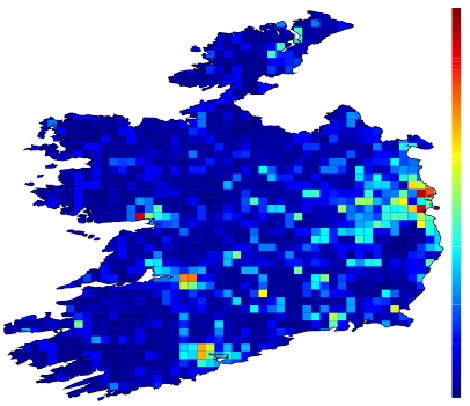}
\includegraphics[width=0.2\textwidth]{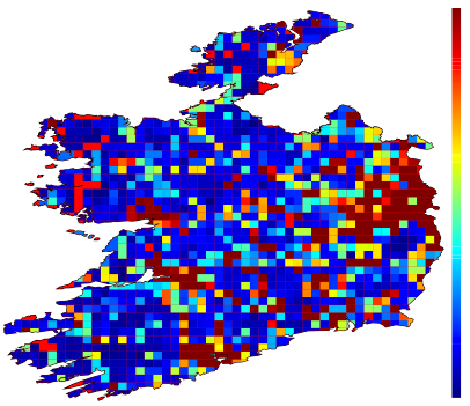}
\caption{
Reference scenario. Blue areas correspond to low demand, red ones to high demand.
The left plot refers to the present time, i.e.,~$k=1$, the right plot to~$k=K=60$
months.
\label{fig:scenario}
}
\end{figure}
We study the performance of our algorithms and the factors affecting it in a large-scale, real-world scenario.
In this section we describe our reference topology and traffic demand, as well as the simulator we employ.

\noindent{\bf Topology}
We leverage two demand and deployment traces, provided by two Irish operators. They consist of two weeks of call-detail records (CDR) information for both voice and data traffic, collected over the whole Republic of Ireland. They include position, (approximate) coverage, and sectorization information for over 6,000 base stations, which constitute our set~$\Bc$. For each base station $b$, the corresponding type~$\Tc(b)$ (e.g., GSM, 3G, LTE) is also given.

\noindent{\bf Traffic demand}
We populate the set~$\Cc$ of subscriber clusters by sampling Ireland's territory
according to the demographic data supplied by the Irish Central Statistics
Office~\cite{census2011} in such a way that each of 
them accounts for (i) at most~$500$ people and (ii) at most~$5~km^2$. By doing this we place around
31,000 subscriber clusters in our topology.
We experimented with different population and area limits, obtaining essentially the same
results we present later.
Traffic demand~$\tau(c,o,1)$ at the present time slot~$k=1$ is obtained
from the traces and summarized in the left plot of \Fig{scenario}.
Future demand is projected according to the Cisco forecast~\cite{cisco}. Our time horizon is~$|\Kc|=60$~time 
periods, with each period representing one month. The total demand for~$k=K=60$, represented on the right-hand 
side of~\Fig{scenario}, is six times the initial one.

\noindent{\bf Simulation and updates}
The sheer scale of our reference topology rules out network simulators such as ns-2 and OMNeT++; rather, we resort to a custom simulator written in Python.
The SINR and throughput between any (base station, subscriber cluster) pair are computed through the OFCOM-vetted
methodology in~\cite{annex-14}, under the assumption of a reuse factor of~$1$.
The fraction $\phi$ of clusters that must enjoy the target competition level is set to $0.7$, unless
otherwise specified.

\begin{figure}
\psfrag{nil}{$t_\emptyset$}
\centering
\includegraphics[width=.45\textwidth]{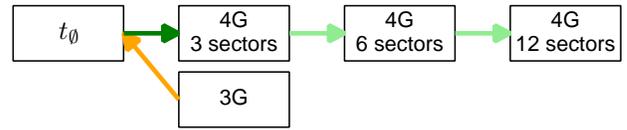}
\caption{
The base station types in~$\Tc$ and the possible changes. 3G base stations can be decommissioned
(orange arrow), i.e., have their type set to~$t_\emptyset$. New LTE base stations can be
created (dark green arrow), possibly in the same location as existing base stations,
or have their capacity enhanced through sectorization (light green arrows).
\label{fig:states}
}
\end{figure}

We assume that the set of base station types~$\Tc$ contains the following elements:
\begin{itemize}
\item the {\em decommissioned} type~$t_\emptyset$;
\item a type for 3G base stations;
\item three types for LTE base stations, with different sectorizations.
\end{itemize}
The changes we can apply to the network are summarized in \Fig{states}: we can {\em decommission}
a 3G base station, or {\em create} a new LTE base station (possibly in the same location of an
existing one), or {\em enhance} the capacity of an existing LTE base station by increasing
the sectorization thereof.
In the following, we will collectively refer to the last two operations as {\em updates}.
It is worth stressing that these limitations are not inherent to our model,
which is able to account for any kind of network update and to interface with any simulator
(see \Fig{simulator}), but merely a way to simplify (and speed up) our performance evaluation.

\section{Results}
\label{sec:results}

\begin{figure*}[t]
\centering
\subfigure[\label{fig:vark-n4}]{
 \includegraphics[width=0.31\textwidth]{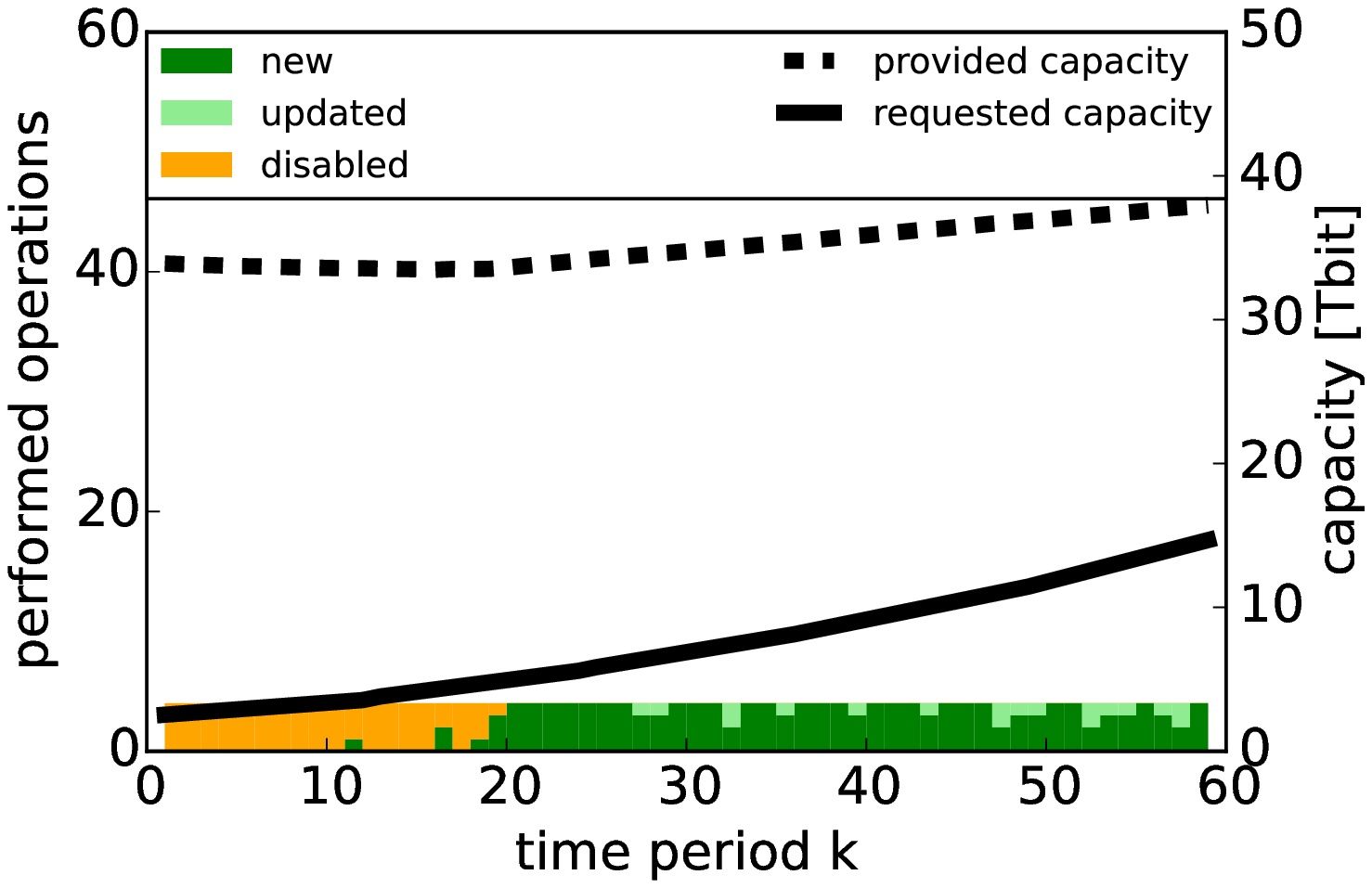}
}
\subfigure[\label{fig:vark-n16}]{
 \includegraphics[width=0.31\textwidth]{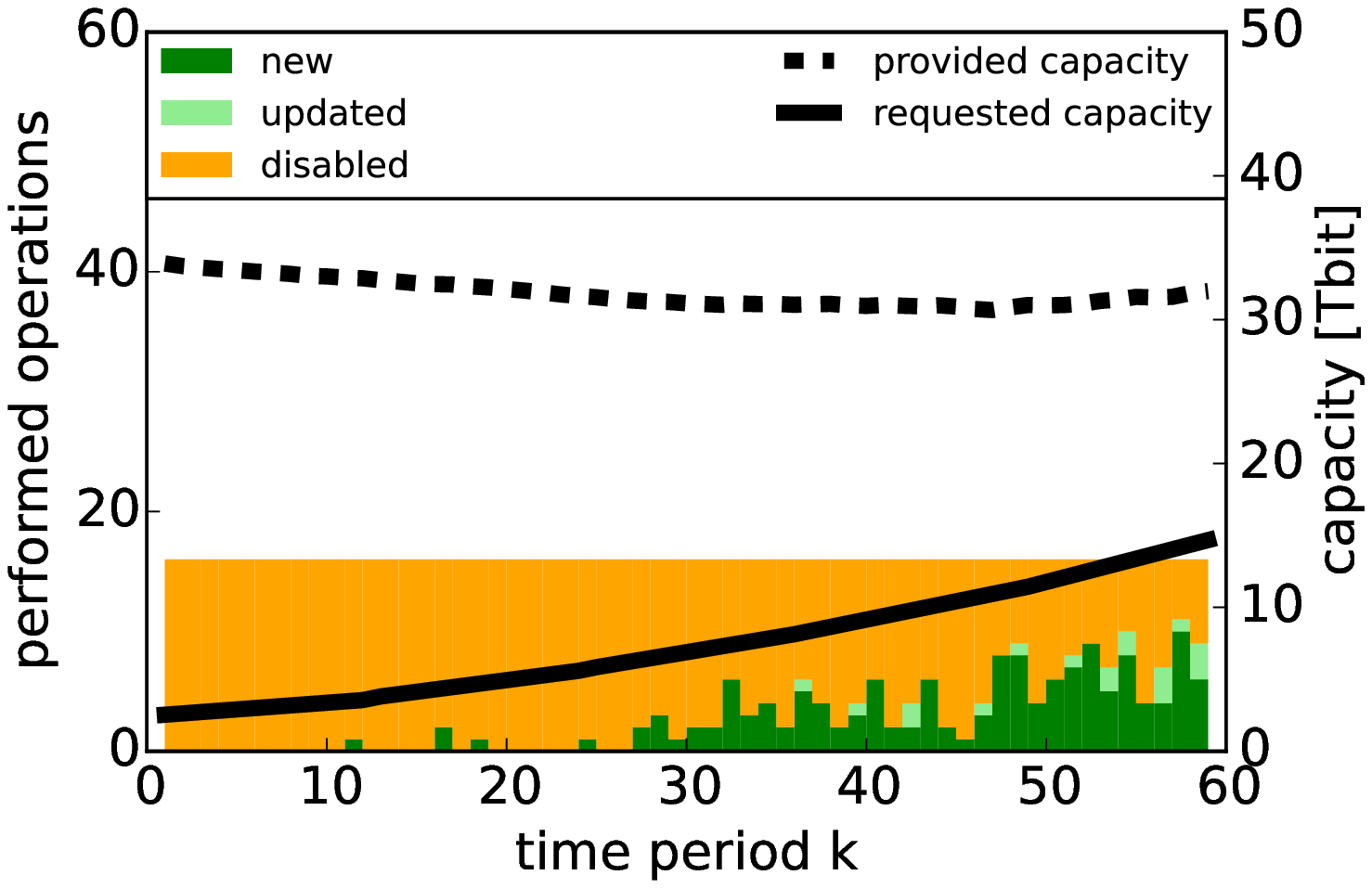}
}
\subfigure[\label{fig:vark-n32}]{
 \includegraphics[width=0.31\textwidth]{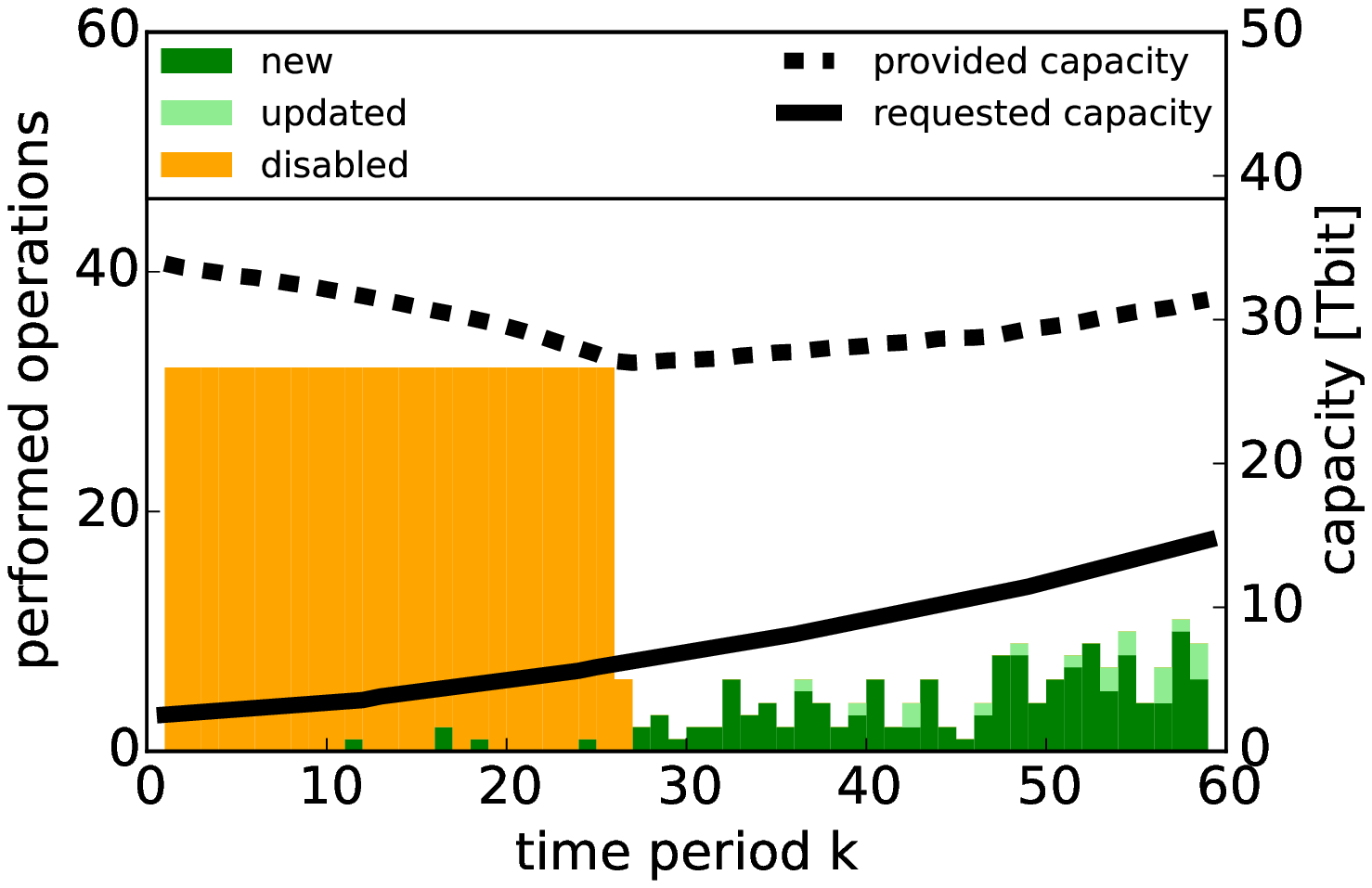}
}
\caption{
Changes applied to the network and requested and provided capacity for each time
period~$k$, when (a)~$N=4~(=N_{\min})$, (b)~$N=16$, (c)~$N=32$.
\label{fig:vark}
}
\end{figure*}

We begin by looking at which network changes are performed at each time period~$k$, and how they impact network capacity.
In \Fig{vark}, solid and dotted lines correspond to requested traffic~$\tau$ and provided capacity~$\sigma$ respectively;
bars represent the number of
created, enhanced and decommissioned base stations.
We are setting in the most favorable case: networks can be operated jointly and there is no competition constraint, i.e.,~$H_{\max}=1$.

\Fig{vark-n4} represents the case for~$N=4$, the minimum possible value of~$N$ in our scenario, as given by \Eq{necessary}.
It is easy to see that the value of~$N$ directly maps to the maximum height of the bars.
Very low values of~$N$, as in \Fig{vark-n4}, imply that most of the changes operators are
able to perform are updates (i.e., create or enhance base stations), so as to meet the demand goal \Eq{goal1}.
As~$N$ increases, as in \Fig{vark-n16}, we are able to decommission more base stations,
and to push forward in time all the updates.
For even larger values of~$N$, we see that there are some time periods where we perform fewer
operations than we could, i.e.,~$\sum_{b\in\Bc,t\in\Tc}x(b,k,t)<N$,
as it happens for~$k>25$ in \Fig{vark-n32}.
This is because we scheduled to decommission all possible base stations at earlier times,
as mandated by \Line{alg3-besttime} in \Alg{disable},
and schedule all needed updates later in time, as in \Line{alg1-besttime} of \Alg{capacity}.

Looking at provided and requested capacity, we can notice that the provided capacity is always substantially higher
than the demand.
This is because we have to preserve the coverage
in the entire topology, i.e., all subscriber clusters~$c\in\Cc$.
In sparsely populated areas, this inevitably translates into underutilized base stations
that operators have no way to decommission.
It is also interesting to see how~$N$ influences the evolution of provided capacity:
low values of~$N$ imply that the capacity slowly increases as updates are performed
(\Fig{vark-n4}).
Higher values of~$N$, as in \Fig{vark-n16}, mean that we can observe the behavior we were
expecting in \Fig{concept}, with network capacity first slowly decreasing due to decommissioning base
stations and then leveling up due to the concurrent scheduling of updates and decommissions.
As we can see from \Fig{vark-n32}, further increasing~$N$ implies that network capacity decreases
more swiftly (as operators can be quicker at decommissioning base stations) and increases more quickly
afterwards, as most updates take place.
Both effects are consistent with our intuition and expectations (\Fig{concept}): being able to perform more changes to the network means being more effective in tracking the traffic demand.

\begin{figure*}[t]
\centering
\subfigure[\label{fig:share-varn}]{
 \includegraphics[width=0.31\textwidth]{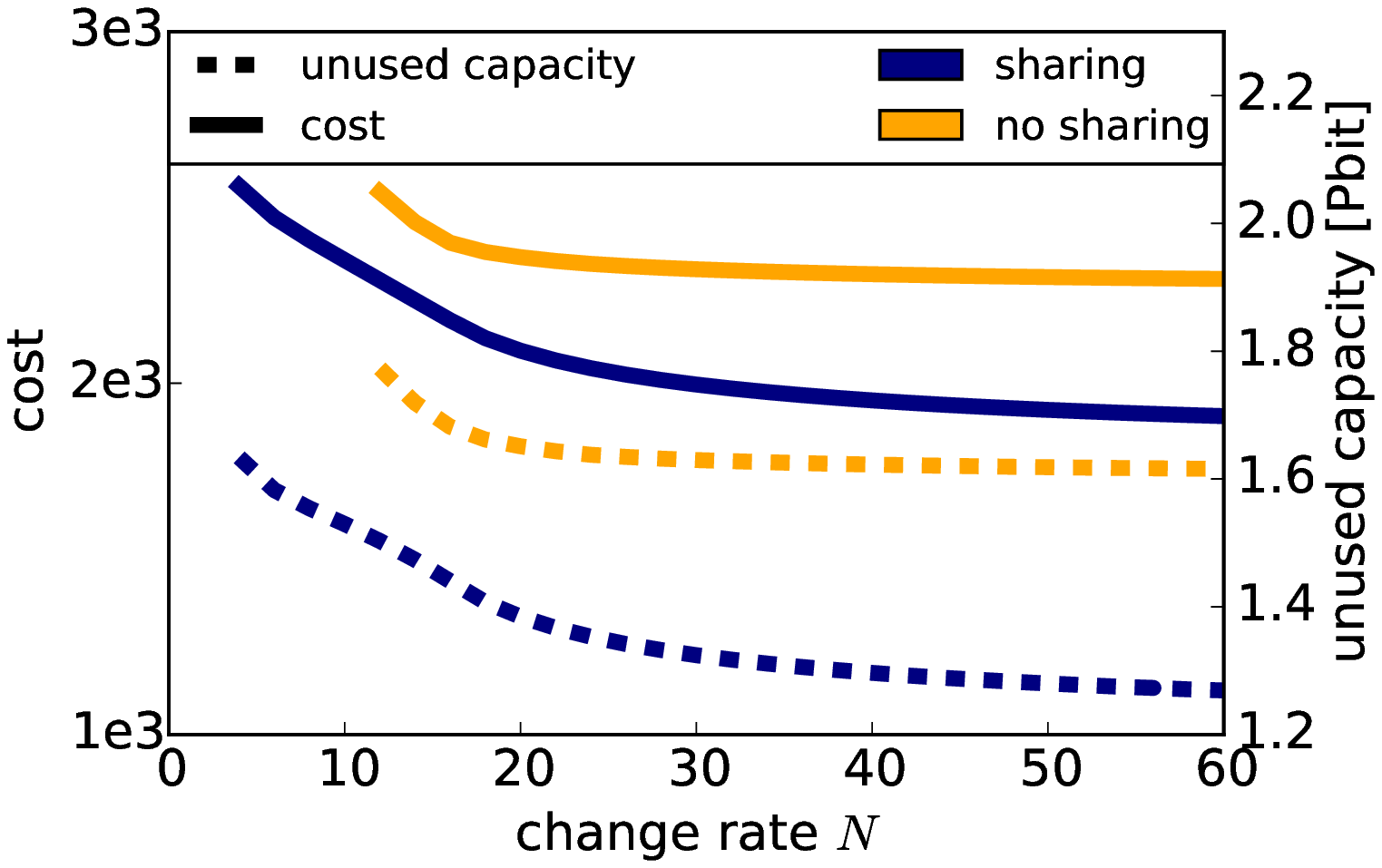}
}
\subfigure[\label{fig:share-map-noshare}]{
 \includegraphics[width=0.31\textwidth]{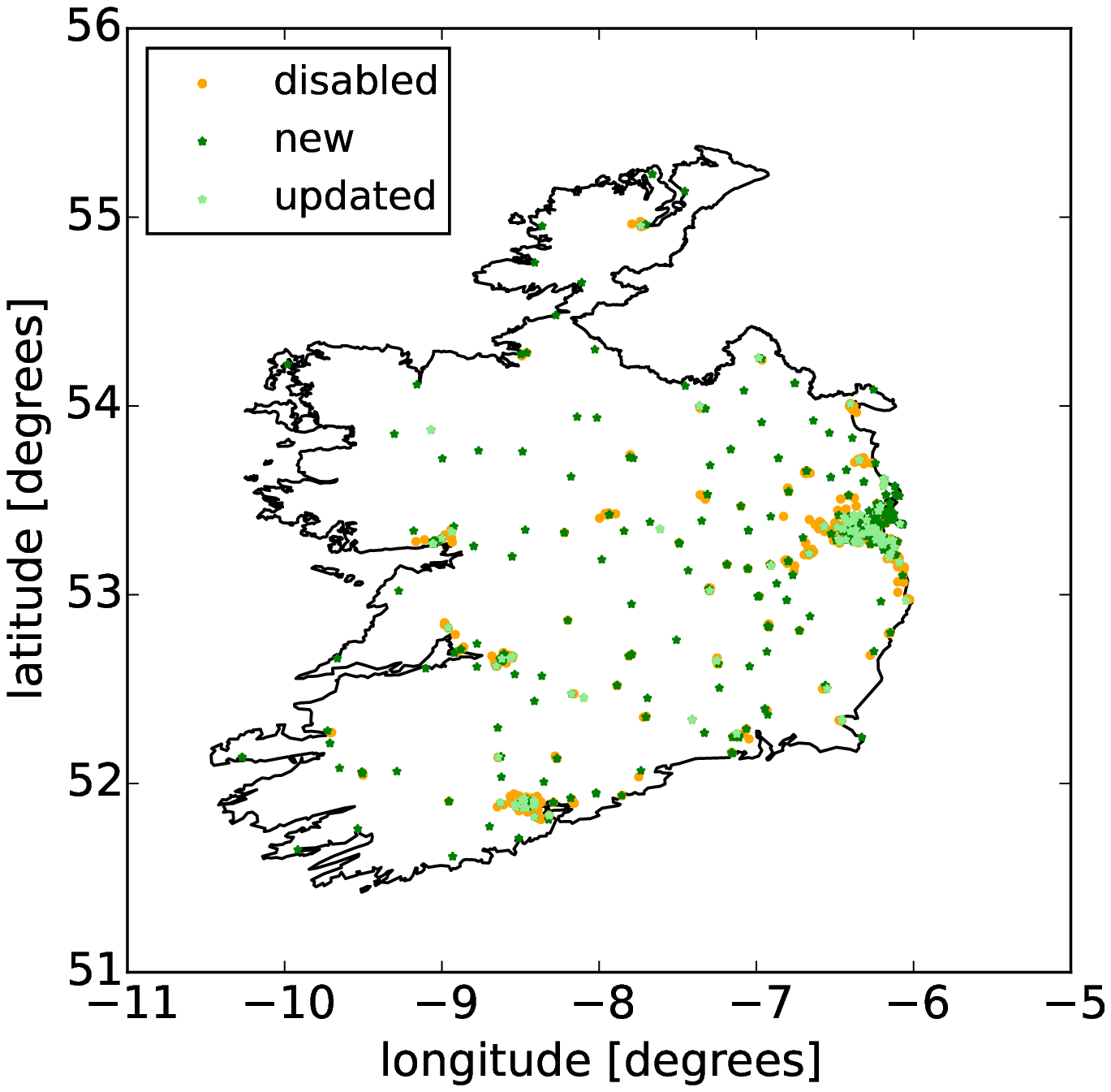}
}
\subfigure[\label{fig:share-map-share}]{
 \includegraphics[width=0.31\textwidth]{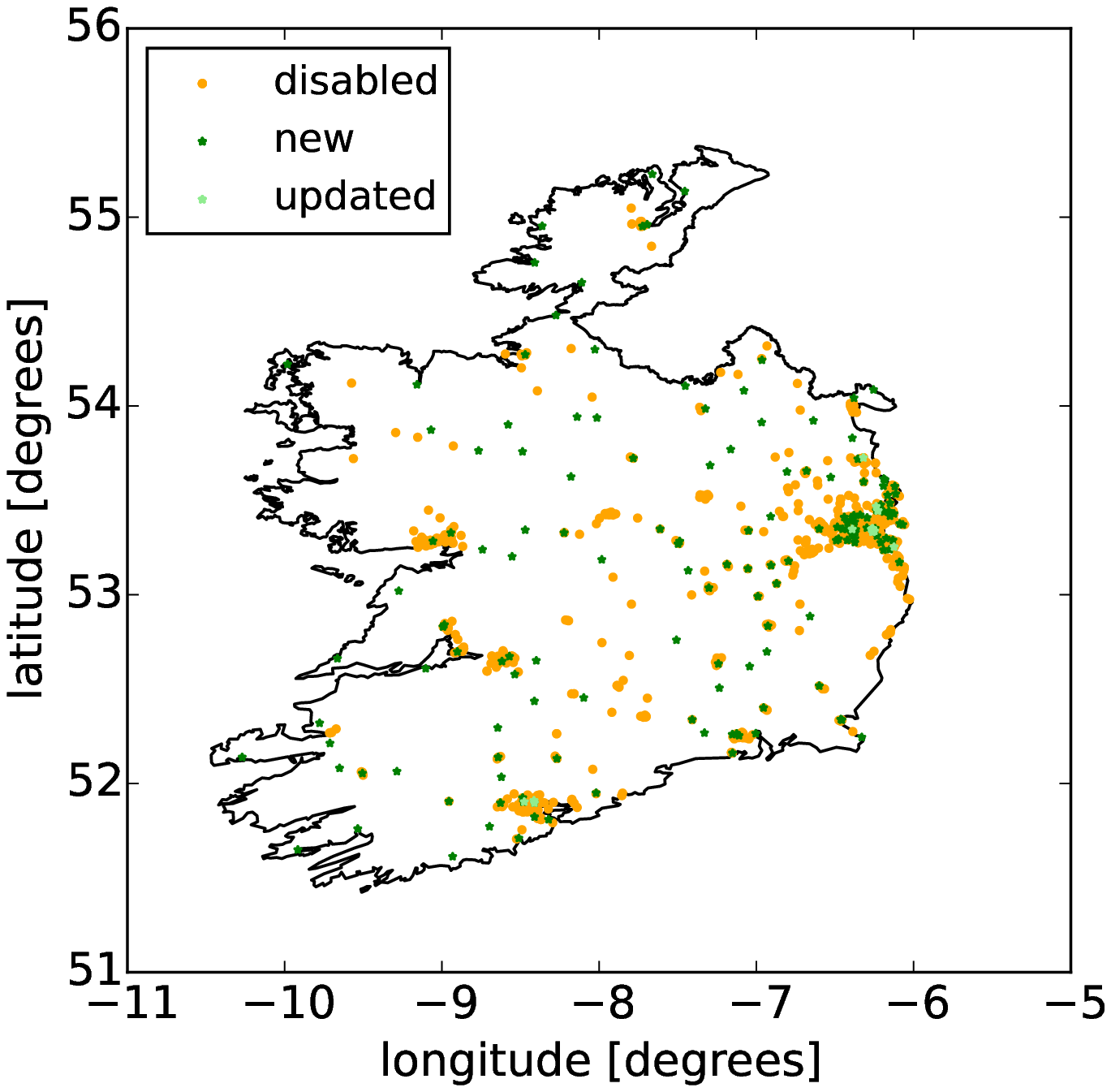}
}
\caption{
Unused capacity and cost savings (as defined in \Eq{goal3})
as a function of~$N$ with and without sharing (a);
location of updated and decommissioned base stations without (b) and with (c) sharing.
\label{fig:share}
}
\end{figure*}

In \Fig{share} we look at the benefits of sharing, i.e., what savings
operators can obtain by operating and updating their networks in a shared fashion.
\Fig{share-varn} is fairly clear -- the benefits of sharing are very significant.
The main effect of allowing sharing is that operators can save substantially more on operational costs,
i.e., decommissioning more base stations.
Sharing also reduces the unused capacity, which however remains quite significant, due to coverage requirements,
as we already observed from \Fig{vark}.

The maps in \Fig{share-map-noshare} and \Fig{share-map-share} show {\em where} base stations are
updated and decommissioned.
Focusing on \Fig{share-map-noshare}, which refers to the case where no sharing
is allowed, we can observe that most updates are concentrated in densely populated areas
(e.g., Dublin in the East), but some take place also in rural areas.
On the other hand, virtually all decommissioned base stations are located in rural and
suburban areas. 
Allowing sharing and moving to \Fig{share-map-share}, we see a very different picture.
In rural areas operators have to update much fewer base stations and can decommission many more;
furthermore,
many base stations can be decommissioned also in urban areas, e.g., in Dublin.

Put together, these results confirm the intuitive notion that network sharing directly translates into better network efficiency.
Backed by our real-world demand and deployment traces, we can add that said efficiency is mostly attained by decommissioning underutilized base stations and, to a lesser extent, by pooling updated ones.
Location-wise, we can say that network updates have the overall effect of {\em migrating} capacity from rural areas to urban ones, and sharing makes such an effect more pronounced, taking advantage of the redundancies of deployments, in particular in the rural areas.

\begin{figure*}[t]
\centering
\subfigure[\label{fig:competition-varh}]{
 \includegraphics[width=0.4\textwidth]{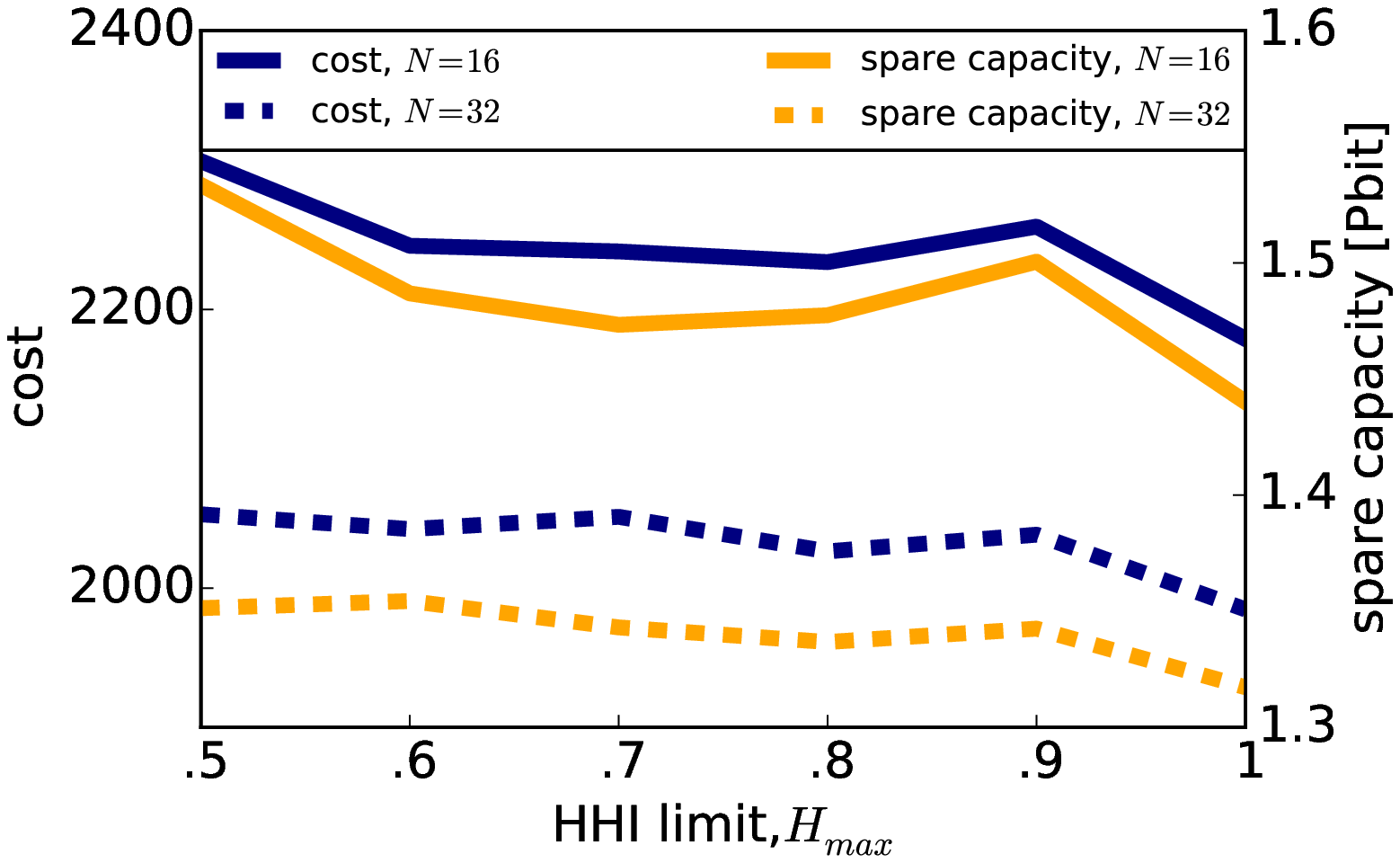}
}
\subfigure[\label{fig:competition-map}]{
 \includegraphics[width=0.4\textwidth]{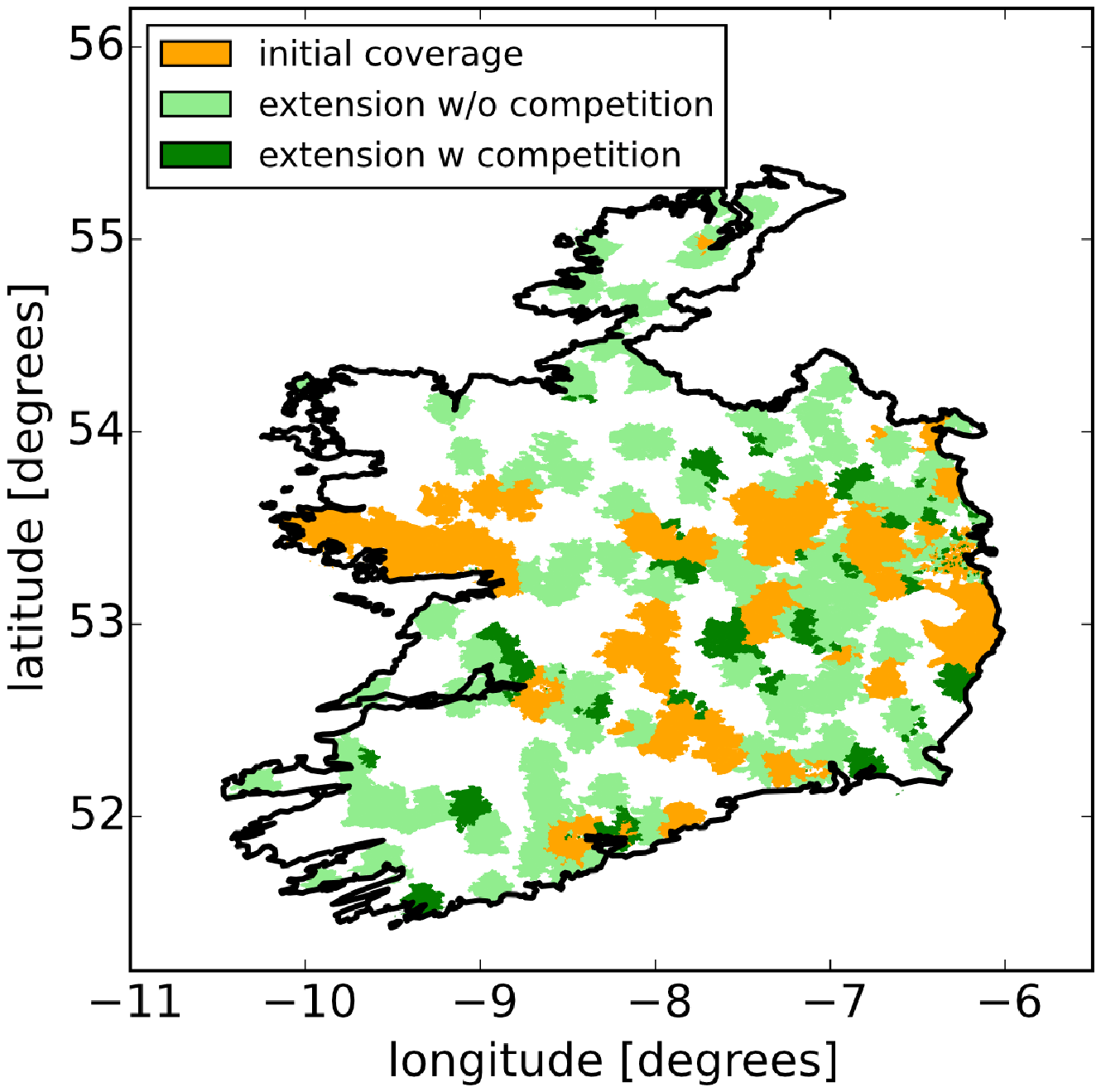}
}
\caption{
Unused capacity and cost (as defined in \Eq{goal3})
as a function of HHI index and for different values of~$N$ (a);
LTE coverage at~$k=0$ (orange),~$k=K=60$ when~$H_{\max}=1$ (light green),
and at~$k=K=60$ when~$H_{\max}=0.5$ (dark green).
\label{fig:competition}
}
\end{figure*}

Competition regulation brings its own requirements on network planning, in particular mandating a certain level
of extra capacity which could be used by a potential new virtual provider.
In \Fig{competition}, we investigate the effects of competition regulation on the evolution of the network infrastructure.
Recall that~$H_{\max}=1$ means that there is no regulation in place, while~$H_{\max}=0.5$ corresponds to the most stringent regulation, imposing as much as~50\% idle capacity.

\Fig{competition-varh} confirms that the tighter the competition regulation,
the lower the savings operators are able to achieve.
As expected, moving from the most loose to the tightest level of regulation
also increases the unused capacity -- i.e., from the regulator's viewpoint, the capacity
available to new operators.
In fact, imposing a minimum competition level has a double effect:
first, it forces the operators to perform 
some updates that otherwise would have not been necessary to meet the capacity goal in \Eq{goal1};
second, it restrains the operators from decommissioning some underutilized base stations.
Intuitively, since most of the decommissioning would take place in the rural areas and most of the updates
in cities (\Fig{share-map-share}), the regulation has the effect that
users from both sparsely and densely populated areas
are able to choose between more (actual or potential) operators.

\Fig{competition-map} gives us further insights on the effect of
competition regulation on LTE coverage.
Specifically, it shows which areas are covered by LTE in the original deployment (i.e.,~$k=0$),
which additional areas are covered at~$k=K=60$ if~$H_{\max}=1$, i.e., no competition regulation
is in place, and which additional areas are covered at~$k=K=60$ if~$H_{\max}=0.5$.
By forcing the installation of more base stations throughout
different areas of the country,
the regulators are also stimulating the deployment of newest technology in areas that otherwise
the operators would not find attractive.

\begin{figure*}
\centering
\subfigure[\label{fig:no-competition-traffic}]{
 \includegraphics[width=0.4\textwidth]{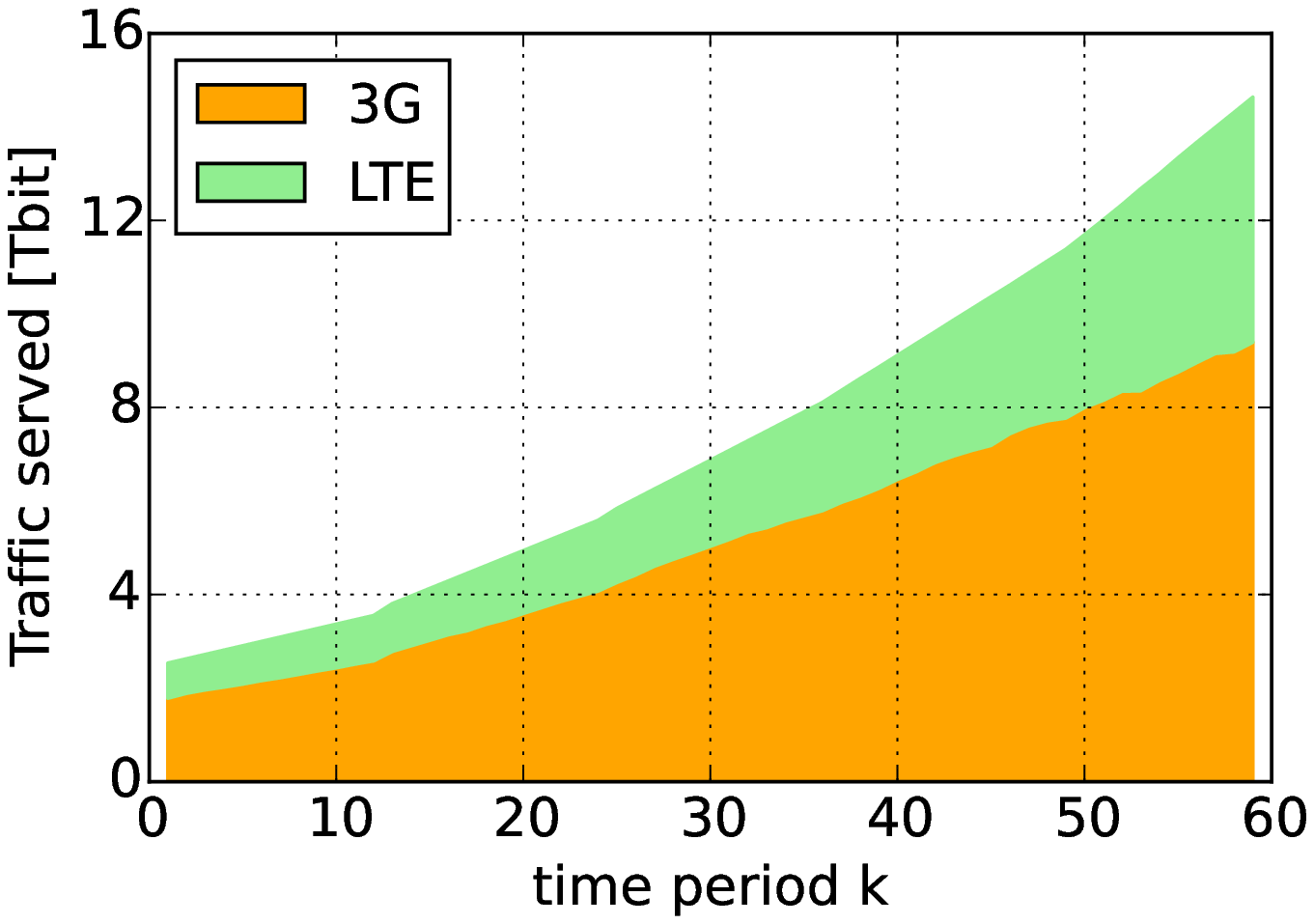}
}
\subfigure[\label{fig:competition-traffic-high}]{
 \includegraphics[width=0.4\textwidth]{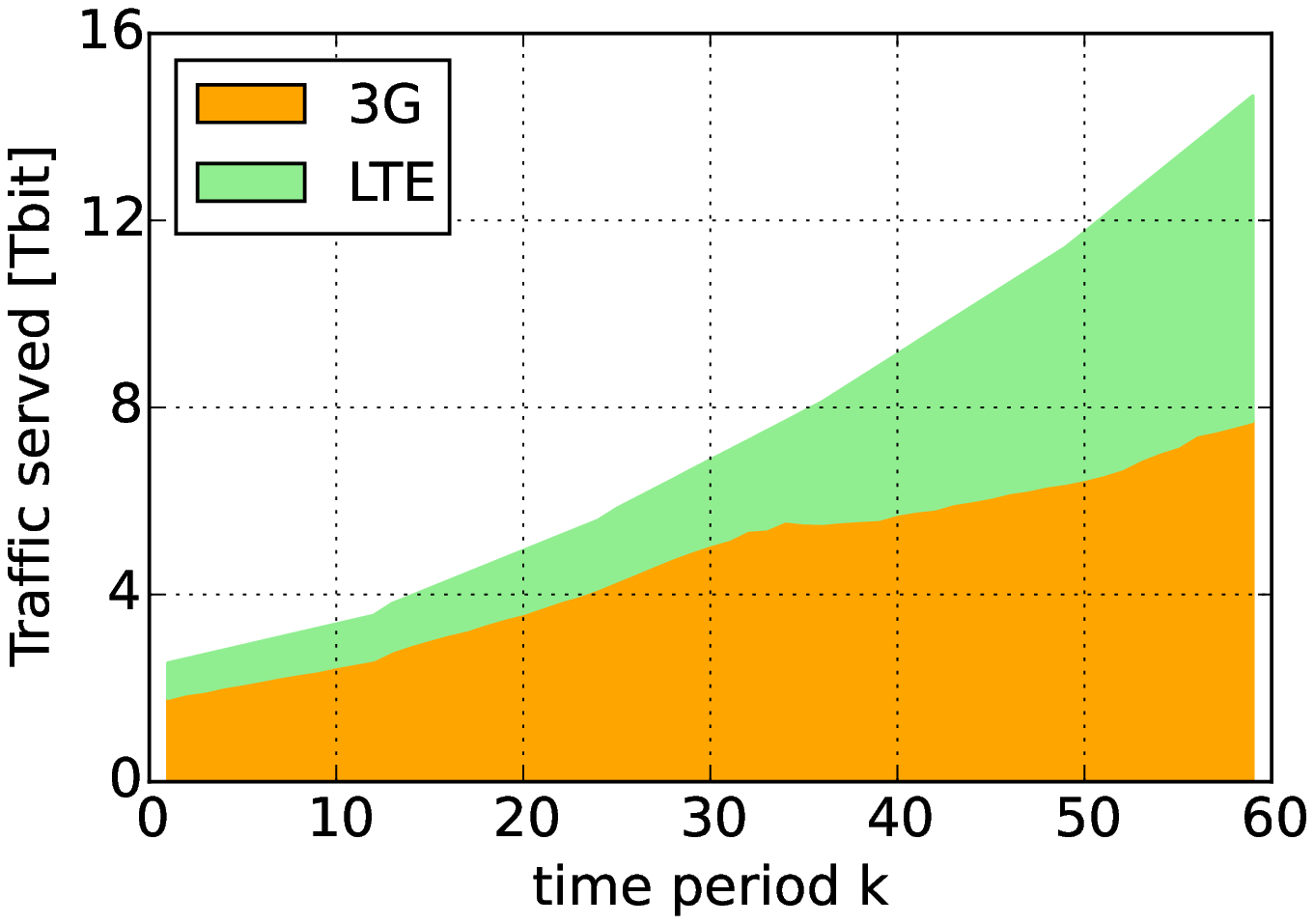}
}
\label{fig:competition-traffic}
\caption{Traffic served per technology for each time period~$k$, when $H_{\max}=1$ (a), and $H_{\max}=0.5$ (b).}
\end{figure*}

As newer technologies are deployed throughout the country, while the older ones tend to be dismissed,
it is interesting to see if the traffic follows the same trend.
In \Fig{competition-traffic} we consider a moderate change rate ($N=16$) and different levels of competition,
and present how much traffic will be served through 3G and 4G. A first observation we can
make concerns the role of 3G -- it will always serve a substantial share of the demand.
Remarkably, this is exactly what the Cisco VNI foresees~\cite{cisco}.

Increasing the level of competition, i.e., going from \Fig{no-competition-traffic}
to \Fig{competition-traffic-high} has the effect of increasing the share of 4G traffic.
This is essentially due to the fact that more 4G infrastructure is deployed, and users tend
to prefer 4G over 3G when they can choose.

As a result, our model is able to capture the tradeoff between \emph{savings} and \emph{the promotion of innovations} which is one of the main goals of regulators in the telecommunication industry. 

\section{Related work}
\label{sec:related}

Our work studies the effects of infrastructure sharing and competition regulation on cellular network planning.

The classic network planning problem usually considers a single operator that aims at minimizing 
operational costs for a specific technology (e.g. 3G, LTE)
while maintaining acceptable users satisfaction both in terms of coverage and capacity.
The research following this line is vast and includes various aspects. For example, in 
\cite{amaldi2003,lee2000,khalek2011,shangyun2010,gordejuela2009,guo2013,ghazzai2015} the authors study the 
optimization of base station location in an area of interest.
Some of these works~\cite{amaldi2003,lee2000,khalek2011,shangyun2010} deal with 3G systems and they
are based on meta-heuristics aiming at minimizing the number of base stations to be deployed. Other 
more recent works~\cite{gordejuela2009,guo2013,ghazzai2015} have focused on the same objective using similar 
approaches but on LTE networks.The work in~\cite{guo2013} in particular uses a model based on stochastic geometry, 
and the coverage probability as the metric to optimize.
The problem addressed in our paper is fundamentally different from the ones addressed by the aforementioned works 
since it includes sharing of already existing infrastructure by more than one operator and considers 
the competition regulation impact, modeled similarly as imposed by the Irish regulator recently~\cite{vodafone-merger}, on network planning.

Resource sharing in inter-operator cellular networks, in fact, is another important aspect of our work and it has been studied in~\cite{panchal2013,hua2012,difrancesco2014,kibilda2013}.
In~\cite{panchal2013} the authors analyze feasible sharing options in the near-term in LTE using co-located and non co-located base stations.
Authors in~\cite{hua2012} assess the benefit of sharing both infrastructure and spectrum, using real base station deployment data.
In~\cite{difrancesco2014} the authors have studied sharing opportunities between two operators, by looking at 
spatial variation in demand peaks.
All the results obtained in these works suggest that resource sharing, whether spectrum or infrastructure,
increases the network capacity and the ability to satisfy users' requirements.
However none of them addresses the problem of \emph{how} to efficiently plan a shared network composed 
by resources already deployed by existing operators.
In our previous work~\cite{kibilda2013} we investigate the \emph{coverage efficiency} obtained by combining 
existing cellular networks considering the coverage redundancy in real deployments in Poland but, unlike this 
paper, our earlier work does not address capacity requirements nor regulatory constraints.

Demographic data are an important source of information for operators to make planning decisions.
Combining such data with the traffic demand information at their base stations, operators can have a clearer 
picture of the needed planning interventions.
Researchers rarely use real topologies and demographic data; they typically rely on simplified synthetic 
topologies often featuring regular, lattice-like deployments that do not reflect the complexity and heterogeneity 
of cellular networks.
Among the studies that uses real data that are related to our work are~\cite{kibilda2013} and~\cite{hua2012}, 
that however do not use traffic demand data.

Other works use traffic demand, either concentrating on profiling the \emph{users}~
\cite{shafiq2012,keralapura2010,paul2011}, or focusing more on the \emph{network} behaviour~
\cite{peng2011,willkomm2008,paul2011,paul2012} but none of them investigates long term planning decisions.
The work in~\cite{peng2011} is of interest to us because it has an objective orthogonal to ours;
it is focused on energy savings and green networking and envisions dynamically switching off base stations at 
off-peak times in different parts of the topology.
Our work instead aims at reconfiguring the whole network, operating long-term (possibly, permanent) changes in 
its infrastructure.
It is important to stress that the two approaches can, and indeed should, coexist: once we reconfigure the 
infrastructure through our network sharing scheme, we can manage it in an energy-efficient fashion.

Works such as~\cite{coucoubertin2012} focus on economic aspects of resource sharing from a network virtualization perspective. It describes the incentives operators have to pool their resources together.
From an implementation perspective, the analysis of cooperative sharing arrangements presented in~\cite{Coopetition} highlights the diversity of approaches currently being used in existing networks, their successes and failures. It points to a process of learning within industry as to which sharing modes allow for both competition and cooperative sharing to thrive. A study of Pakistan's experience of network sharing indicates the varying economic gains made in a still-developing market by adopting different sharing strategies~\cite{pakistan}. 

However, while these papers address some of the economic effects of sharing, none of these has dealt with regulatory concerns regarding the ensuing market concentration which occurs through network sharing in mature mobile markets. Our study is unique because it combines all the aforementioned aspects (real data analysis, spatial distribution of traffic, network sharing, and network planning), and it considers the impact of the limitations imposed by the regulators on the savings when managing two networks in a shared fashion in a mature market.

\section{Conclusion}
\label{sec:conclusion}

We studied the modernization phase cellular networks will go through in the near future:
mobile operators will decommission some underutilized base stations in order to save on costs,
and deploy new-generation base stations
in order to cope with the increasing demand.
Operators can join forces and perform said changes in a shared way, so as to improve the efficiency
of their networks.
Operators' ability to share infrastructure may be constrained by competition regulation, and
the speed with which operators can make changes to their own networks may also be limited by
practical considerations. We incorporate both factors in out study of cellular network planning.

Our first contribution is a
general framework that describes
network modernization scenarios, accounting for real topologies and demand information,
and including multiple base station technologies. This model was presented in \Sec{sysmodel}.
Given our model and a limited budget of changes
to perform at each time period, we presented in \Sec{problem} a family of algorithms
able to schedule the changes in a cost effective manner, while satisfying the demand
and complying with regulatory constraints. These algorithms work unmodified whether operators
perform their updates individually or in a shared fashion and, as shown in \Sec{discussion},
return quasi-optimal solutions with a very small computational complexity,
dominated by their sorting stage.

We apply our algorithms in a large-scale scenario, built from real-world demand and deployment
traces as described in \Sec{scenario}. As summarized in \Sec{results},
we found that network
modernization essentially means moving capacity from rural, sparsely populated
areas (where many base stations can be decommissioned)
to urban ones (where most of the new-generation base stations are located).
Allowing sharing, i.e.,
permitting operators to jointly update and manage their networks, greatly enhances their
effectiveness. Such benefits are reduced if tight competition rules are in place, but
never entirely jeopardized.
Indeed, tight competition regulations have the secondary effect of stimulating operators to
extend the capacity and coverage of their new-generation networks,
which can therefore serve a larger fraction of their demand.

\bibliographystyle{IEEEtran}

\input{arXiv_two_columns.bbl}

\appendices

\clearpage\newpage

\setcounter{lemma}{0}
\setcounter{property}{0}
\section*{Appendix: Proofs}
\begin{property}
Both goal 1 and goal 2 can be reached through network capacity-preserving changes alone, i.e.,
changes that comply with \Eq{preserve}.
\end{property}
\begin{IEEEproof}
Goal 1 means to satisfy \Eq{goal1} for all subscriber clusters~$c\in\Cc$ and time periods~$k\in\Kc$.
If this is not the case, then the solution is scheduling updates that increase capacity.
Decreasing the capacity for some subscriber clusters, i.e., breaking \Eq{preserve}, is never necessary.
A similar but slightly different reasoning holds for goal 2. Increasing
the~$\sigma$-values, as we can see from \Eq{hhi}, decreases the HHI.
\end{IEEEproof}

\begin{property}
If the initial configuration satisfies goal 1, then pursuing goal 2 by scheduling further capacity-preserving
changes does not compromise goal 1.
\end{property}
\begin{IEEEproof}
Once again, let us look at \Eq{goal1}: if it holds, then all~$\sigma$-values are no lower than the corresponding~$\tau$-ones,
therefore, there is no way that further increasing the~$\sigma$-values can change this.
\end{IEEEproof}

\begin{property}
\label{prop:complexity}
The worst-case, {\em combined} time complexity of all our algorithms is~$O(|\Bc|log|\Bc|)$.
\end{property}
\begin{IEEEproof}
At each iteration of each of our algorithms, we make exactly one decision, i.e., set one~$x$-value to~$1$.
Even if each base station is updated once to each possible type, the total number of decisions is still
bounded by~$|\Tc||\Bc|$, under the reasonable assumption that~$|\Bc|\gg|\Tc|$, dominated by the sorting in 
\Line{alg3-sort} in \Alg{disable}, which has complexity~$O(|\Bc|\log|\Bc|)$.
\end{IEEEproof}

\begin{property}
\label{prop:feasible}
If a set of changes satisfies \Eq{necessary}, then \Alg{capacity} is able to schedule it.
\end{property}
\begin{IEEEproof}
Scheduling a set of changes means enacting each of them at time~$\hat{k}\leq k^\star$ no later than its deadline.
In other words, every time we reach \Line{alg1-besttime} in \Alg{capacity},
the set~$\{ h\in\Kc: \sum_{b\in\Bc,t\in\Tc} x(b,h,t) < N\wedge h\leq k^\star \}$ must be non-empty.
In \Line{alg1-mosturgent}, we always select
to schedule the base station with the lowest value of~$k^\star$.
This means that if at
the current iteration we are scheduling a change due at time period~$k^\star$, then all
the changes we scheduled so far were due at~$k^\star$ or earlier.

Since \Eq{necessary}
holds, the number of such changes it at most~$Nk^\star-1$;
therefore, there must be a~$\hat{k}$ between~$1$
and~$k^\star$ for which fewer than~$N$ changes have been scheduled. 
Hence, the set is non-empty.
\end{IEEEproof}

\begin{property}
\label{prop:optimal}
The schedule returned by \Alg{capacity} minimizes the quantity in \Eq{tomin}.
\end{property}
\begin{IEEEproof}
We prove the property by induction.

\noindent{\em Initialization.}
At the first iteration of \Alg{capacity}, all~$x$-values are set to~$0$, thus in \Line{alg1-besttime}
we have~$\hat{k}=k^\star$. The value of the quantity in \Eq{tomin} is zero, hence the schedule is optimal.

\noindent{\em Induction step.}
Suppose all other changes have been scheduled optimally, i.e., they cannot be moved forward in time. \Alg{capacity}
will try (\Line{alg1-besttime}) to schedule the current change for period~$k^\star$, then~$k^\star-1$, and so on,
stopping at the latest feasible time. It follows that the resulting schedule still has the lowest possible value of \Eq{tomin}.
\end{IEEEproof}

\end{document}

%% file: arXiv_two_columns.bbl

%% file: arXiv_two_columns.bbl
\begin{thebibliography}{10}
\providecommand{\url}[1]{#1}
\csname url@samestyle\endcsname
\providecommand{\newblock}{\relax}
\providecommand{\bibinfo}[2]{#2}
\providecommand{\BIBentrySTDinterwordspacing}{\spaceskip=0pt\relax}
\providecommand{\BIBentryALTinterwordstretchfactor}{4}
\providecommand{\BIBentryALTinterwordspacing}{\spaceskip=\fontdimen2\font plus
\BIBentryALTinterwordstretchfactor\fontdimen3\font minus
  \fontdimen4\font\relax}
\providecommand{\BIBforeignlanguage}[2]{{%
\expandafter\ifx\csname l@#1\endcsname\relax
\typeout{** WARNING: IEEEtran.bst: No hyphenation pattern has been}%
\typeout{** loaded for the language `#1'. Using the pattern for}%
\typeout{** the default language instead.}%
\else
\language=\csname l@#1\endcsname
\fi
#2}}
\providecommand{\BIBdecl}{\relax}
\BIBdecl

\bibitem{cisco}
\BIBentryALTinterwordspacing
Cisco, ``{Cisco Visual Networking Index: Global Mobile Data Traffic Forecast
  Update, 2013-2018}.'' [Online]. Available:
  \url{http://www.cisco.com/c/en/us/solutions/collateral/service-provider/visual-networking-index-vni/white_paper_c11-520862.pdf}
\BIBentrySTDinterwordspacing

\bibitem{difrancesco2014}
P.~{Di Francesco}, F.~Malandrino, and L.~A. DaSilva, ``{Mobile Network Sharing
  Between Operators: A Demand Trace-Driven Study},'' in \emph{ACM SIGCOMM
  Capacity Sharing Workshop (CSWS)}, 2014.

\bibitem{leng2014}
B.~Leng, P.~Mansourifard, and B.~Krishnamachari, ``{Microeconomic Analysis of
  Base-station Sharing in Green Cellular Networks},'' in \emph{IEEE
  International Conference on Computer Communications (INFOCOM)}, 2014.

\bibitem{doyle2014}
L.~Doyle, J.~Kibi{\l}da, T.~Forde, and L.~A. DaSilva, ``{Spectrum without
  Bounds, Networks without Borders},'' \emph{Proceedings of the IEEE}, vol.
  102, no.~3, pp. 351--365, 2014.

\bibitem{vodafone-merger}
\BIBentryALTinterwordspacing
{RTE News}, ``{Vodafone criticises Three-O2 merger terms},'' 2014. [Online].
  Available:
  \url{http://www.rte.ie/news/business/2014/0529/620446-vodafone-three-criticism/}
\BIBentrySTDinterwordspacing

\bibitem{green-ajmone}
\BIBentryALTinterwordspacing
M.~A. Marsan and M.~Meo, ``Energy efficient management of two cellular access
  networks,'' \emph{SIGMETRICS Perform. Eval. Rev.}, vol.~37, no.~4, pp.
  69--73, Mar. 2010. [Online]. Available:
  \url{http://doi.acm.org/10.1145/1773394.1773406}
\BIBentrySTDinterwordspacing

\bibitem{green-on}
E.~Oh, B.~Krishnamachari, X.~Liu, and Z.~Niu, ``Toward dynamic energy-efficient
  operation of cellular network infrastructure,'' \emph{Communications
  Magazine, IEEE}, vol.~49, no.~6, pp. 56--61, June 2011.

\bibitem{peng2011}
C.~Peng, S.-B. Lee, S.~Lu, H.~Luo, and H.~Li, ``{Traffic-driven power saving in
  operational 3G cellular networks},'' in \emph{ACM International Conference on
  Mobile Computing and Networking (MobiCom)}, 2011.

\bibitem{mergers}
{U.S. Department of Justice and the Federal Trade Commission}, ``Horizontal
  merger guidelines,'' Tech. Rep., August 2010.

\bibitem{EULaw}
{European Commission}, ``{EU Competition Law Rules Applicable to Merger Control
  Situation as at 1 April 2010},'' Tech. Rep., 2010.

\bibitem{pakistan}
M.~Choudhary, H.~Babar, H.~Shakeel, and A.~Abbas, ``Economics of network
  sharing - a case study of mobile telecom sector in pakistan,'' in
  \emph{Collaborative Computing: Networking, Applications and Worksharing,
  2009. CollaborateCom 2009. 5th International Conference on}, Nov 2009, pp.
  1--6.

\bibitem{inapprox}
\BIBentryALTinterwordspacing
N.~Alon, D.~Moshkovitz, and S.~Safra, ``Algorithmic construction of sets for
  k-restrictions,'' \emph{ACM Trans. Algorithms}, vol.~2, no.~2, pp. 153--177,
  Apr. 2006. [Online]. Available:
  \url{http://doi.acm.org/10.1145/1150334.1150336}
\BIBentrySTDinterwordspacing

\bibitem{census2011}
\BIBentryALTinterwordspacing
 [Online]. Available: \url{http://www.cso.ie/en/census/census2011
  boundaryfiles/}
\BIBentrySTDinterwordspacing

\bibitem{annex-14}
\BIBentryALTinterwordspacing
``{LTE Technical Modelling Revised Methodology},'' \emph{OFCOM White Paper},
  2012. [Online]. Available:
  \url{http://stakeholders.ofcom.org.uk/binaries/consultations/award-800mhz/annexes/annex14.pdf}
\BIBentrySTDinterwordspacing

\bibitem{amaldi2003}
E.~Amaldi, A.~Capone, and F.~Malucelli, ``{Planning UMTS Base Station Location:
  Optimization Models with Power Control and Algorithms},'' \emph{IEEE
  Transactions on Wireless Communications}, vol.~2, no.~5, 2003.

\bibitem{lee2000}
C.~Lee and H.~Kang, ``{Cell Planning with Capacity Expansion in Mobile
  Communications: a TABU Search Approach},'' \emph{IEEE Transactions on
  Vehicular Technology}, vol.~49, no.~5, 2000.

\bibitem{khalek2011}
A.~Abdel-Khalek, L.~Al-Janj, and Z.~Dawy, ``{Optimization Models and Algorithms
  for Joint Uplink/Downlink UMTS Radio Network Planning with SIR-Based Power
  Control},'' \emph{IEEE Transactions on Vehicular Technology}, vol.~60, no.~4,
  2011.

\bibitem{shangyun2010}
L.~Shangyun and M.~St-Hilaire, ``{A Genetic Algorithm for the Global Planning
  Problem of UMTS Netowrks},'' in \emph{IEEE Global Communications Conference
  (GLOBECOM)}, 2010.

\bibitem{gordejuela2009}
F.~Gordejuela-Sanchez and J.~Zhang, ``{LTE Access Network Planning adn
  Optimization: A Service Oriented and Technology-Specific Prospective},'' in
  \emph{IEEE Global Communications Conference (GLOBECOM)}, 2009.

\bibitem{guo2013}
A.~Guo and M.~Haenggi, ``{Spatial Stochastic Models and Metrics for the
  Structure of Base Stations in Cellular Networks},'' \emph{IEEE Transactions
  on Wireless Communications}, vol.~12, no.~11, 2013.

\bibitem{ghazzai2015}
H.~Ghazzai, E.~Yaacoub, M.~S. Alouini, Z.~Dawy, and A.~Abu-Dayya, ``{Optimized
  LTE Cell Planning with Varying Spatial and Temporal User Densities},''
  \emph{IEEE Transactions on Vehicular Technology}, vol.~PP, no.~99, 2015.

\bibitem{panchal2013}
J.~S. Panchal, R.~D. Yates, and M.~M. Buddhikot, ``{Mobile Network Resource
  Sharing Options: Performance Comparisons},'' \emph{IEEE Transactions on
  Wireless Communications}, vol.~12, no.~9, pp. 4470--4482, 2013.

\bibitem{hua2012}
S.~Hua, P.~Liu, and S.~S. Panwar, ``{The Urge to Merge: When Cellular Service
  Providers Pool Capacity},'' in \emph{IEEE International Conference on
  Communications (ICC)}, 2012.

\bibitem{kibilda2013}
J.~Kibi{\l}da and L.~A. DaSilva, ``{Efficient Coverage through Inter-operator
  Infrastructure Sharing in Mobile Networks},'' in \emph{Wireless Days}, 2013.

\bibitem{shafiq2012}
M.~Z. Shafiq, L.~Ji, A.~Liu, J.~J.~Pang, and J.~Wang, ``{Characterizing
  Geospatial Dynamics of Application Usage in a 3G Cellular Data Network},'' in
  \emph{IEEE International Conference on Computer Communications (INFOCOM)},
  2012.

\bibitem{keralapura2010}
R.~Keralapura, A.~Nucci, Z.~L. Zhang, and L.~Gao, ``{Profiling users in a 3G
  network using hourglass co-clustering},'' in \emph{ACM International
  Conference on Mobile Computing and Networking (MobiCom)}, 2011.

\bibitem{paul2011}
U.~Paul, A.~Subramanian, M.~Buddhikot, and S.~Das, ``{Understanding Traffic
  Dynamics in Cellular Data Networks},'' in \emph{IEEE International Conference
  on Computer Communications (INFOCOM)}, 2011.

\bibitem{willkomm2008}
D.~Willkomm, S.~Machiraju, J.~Bolot, and A.~Wolisz, ``{Primary Users in
  Cellular Networks: A Large-Scale Measurement Study},'' in \emph{IEEE
  Symposium on New Frontiers in Dynamic Spectrum Access Networks (DySPAN)},
  2008.

\bibitem{paul2012}
U.~Paul, A.~Subramanian, M.~Buddhikot, and S.~Das, ``{Understanding Spatial
  Relationships in Resource Usage in Cellular Data Networks},'' in \emph{IEEE
  International Workshop on Network Science for Communication Networks
  (NetSciCom)}, 2012.

\bibitem{coucoubertin2012}
C.~Courcoubetis and R.~Weber, ``{Economic Issue in Shared Infrastructures},''
  \emph{IEEE/ACM Transactions on Networking}, vol.~20, no.~2, pp. 594--608,
  2012.

\bibitem{Coopetition}
J.~Markendahl and B.~G. Molleryd, ``{On co-opetition between mobile network
  operators: Why and how competitors cooperate},'' International
  Telecommunications Society (ITS), 19th ITS Biennial Conference, Bangkok 2012:
  Moving Forward with Future Technologies - Opening a Platform for All 72491,
  2012.

\end{thebibliography}
